\documentclass[a4paper,aps,pra,twocolumn,preprintnumbers,amsmath,amssymb]{revtex4}
\usepackage{bbm}
\usepackage{amsmath}
\usepackage{verbatim}
\usepackage{graphicx}
\bibliography{References}
\begin{document}
\newcommand{\avg}[1]{\langle #1 \rangle}

\newcommand{\var}{\textrm{var}}

\title{Robust entanglement generation by reservoir engineering}

\author{Christine A. Muschik$^{1}$, Hanna Krauter$^{2}$, Kasper Jensen$^{2}$, Jonas M. Petersen$^{2}$, J. Ignacio Cirac$^{3}$, and Eugene S. Polzik$^{2}$}

\affiliation{$^{1}$ ICFO-Institut de Ci\`{e}ncies Fot\`{o}niques,
Mediterranean
Technology Park, 08860 Castelldefels (Barcelona), Spain.\\
$^{2}$ Niels Bohr Institute, Danish Quantum Optics Center QUANTOP,
Copenhagen University, Blegdamsvej 17, 2100 Copenhagen, Denmark.\\
$^{3}$ Max-Planck-Institut f\"ur Quantenoptik,
Hans-Kopfermann-Str. 1, D-85748 Garching, Germany.}

\begin{abstract}
Following a recent proposal [C. Muschik et. al., Phys. Rev. A
\textbf{83}, 052312 (2011)], engineered dissipative processes have
been used for the generation of stable entanglement between two
macroscopic atomic ensembles at room temperature [H. Krauter et.
al., Phys. Rev. Lett. \textbf{107}, 080503 (2011)]. This
experiment included the preparation of entangled states which are
continuously available during a time interval of one hour. Here,
we present additional material, further-reaching data and an
extension of the theory developed in [C. Muschik et. al., Phys.
Rev. A \textbf{83}, 052312 (2011)]. In particular, we show how the
combination of the entangling dissipative mechanism with
measurements can give rise to a substantial improvement of the
generated entanglement in the presence of noise.
\end{abstract}


\maketitle 
\section{Introduction}\label{Sec:Introduction}
In a recent experiment \cite{EbDExperiment, EbDExperimentArXiv}, a
new technique for generating extremely robust and long-lived
entanglement has been demonstrated following a  proposal put
forward in \cite{EbDtheory} (see also~\cite{PaSC05}). By means of
reservoir engineering
\cite{PCZ96,PZ98,PHBK99,BBTK00,PH02,KC04,GRSZ06,DEPC07,DMK08,KBD08,VWC09,BPP10,ADK10,KNPM10,LiFicek10,PCC11,CBK11,BMS11,QZK11,KRS11,KBGKE11,KDJ11},
entanglement has been produced purely dissipatively. Moreover, it
has been shown how engineered dissipative processes in combination
with continuous measurements can be used to create entanglement in
a steady state. Using this method, entanglement between two
macroscopic atomic ensembles~\cite{FootnoteMacroscopicity} has
been maintained and verified for up to one hour. This extends the
time intervals during which event-ready entanglement of material
objects can be provided by several orders of magnitude.
In this article, we give an extended description of this
experiment and present additional supporting data. A detailed and
rigorous derivation of theoretical framework of the employed
dissipative entangling mechanism can be found in \cite{EbDtheory}.
Here, we also discuss an extension of this method which has been
used in the experiment and includes measurements. We show how they
can be used to improve the purely dissipative protocol and explain
the basic working principle in detail.\\

The coupling of a quantum system to its environment, commonly
referred to as dissipation, is traditionally considered to be a
main problem impairing experiments involving quantum superposition
states and the development of quantum technologies. Harnessing
dissipative processes rather than aiming for eliminating their
influence is a radically new concept and represents a paradigm
shift in Quantum Information Science. We show that even limited
control of the coupling between system and environment can enable
one to turn a major problem into an asset. This change in
perspective is not only of conceptual interest, but yields also
significant practical advantages. The protocol discussed here
relies on engineered dissipation. More specifically, the coupling
of a system with a reservoir is tailored such that the desired
state is obtained as the steady state of the dissipative
evolution. This way, the target state is reached independently of
the initial conditions. Accordingly, this type of protocol does
not require the precise initialization of the system in a well
defined state. The resulting quantum state can be maintained for
long times since it is stabilized by the dissipative dynamics.
This mechanism continuously drives the system into the desired
entangled state, even in the presence of noise sources, which
limit the coherence time of the quantum system. Thus, the use of
engineered dissipation enables the realization of unlimited
entanglement lifetimes, which is not achievable by traditional
methods.

We consider here two macroscopic atomic ensembles at room
temperature interacting with freely propagating coherent light.
Quantum information can be encoded in collective atomic spin
states which are unaffected by the thermal motion of the atoms.
This system has been shown to provide an excellent platform for
quantum memory schemes and the realization of light-matter
interfaces~\cite{DuCZP00,JSC04,ShSFMP05,Quint,SKO06,HSP10,JWK10,MKHP11}.
In the protocol discussed here, two atomic ensembles are entangled
by virtue of a dissipative mechanism which is induced by the
application of a strong driving field. The generated entanglement
can be accessed at any moment during an extended period of time,
which makes it particularly useful for protocols, where it is not
known in advance when the entangled state is needed  (for example
if probabilistic subroutines are involved). If the entangled state
is to be used, the driving field inducing the entangling mechanism
is switched off before the actual protocol is run. Since
entanglement is created between two ensembles, the resulting
atomic state can either be used directly or be read out on demand
using light-matter interface schemes~\cite{QuantumMemory,FiSOP05}.
Another very interesting application is the use in continuous
protocols, for example in dissipative quantum repeater
schemes~\cite{VMC10}. This type of scheme requires continuous
entanglement for establishing high-quality steady state
entanglement over large distances.

The remainder of the article is organized as follows. In
Sec.~\ref{Sec:Overview}, we summarize the main results and explain
the key features of the scheme.
Sec.~\ref{Subsec:PurelyDissipativeEntanglement} is concerned with
the creation of purely dissipative entanglement. We offer an
intuitive explanation and data supporting this interpretation. We
also compare the basic working mechanism to standard approaches
and highlight the distinguishing features to the dissipative
scheme discussed here. Thereafter, a hybrid method is described,
where the dissipative mechanism is combined with continuous
measurements on the light field. In
Sec.~\ref{Sec:EntanglementAssistedByMeasurements}, we explain how
monitoring of the scattered photons can lead to an improvement of
the produced entanglement in the presence of noise. In
Sec.~\ref{Sec:ExperimentalDetails}, we present additional
experimental material and Sec.~\ref{Sec:Conclusions} concludes the
paper.
\section{Overview and central results}\label{Sec:Overview}
\begin{figure}
\centering \includegraphics[width=8.5cm]{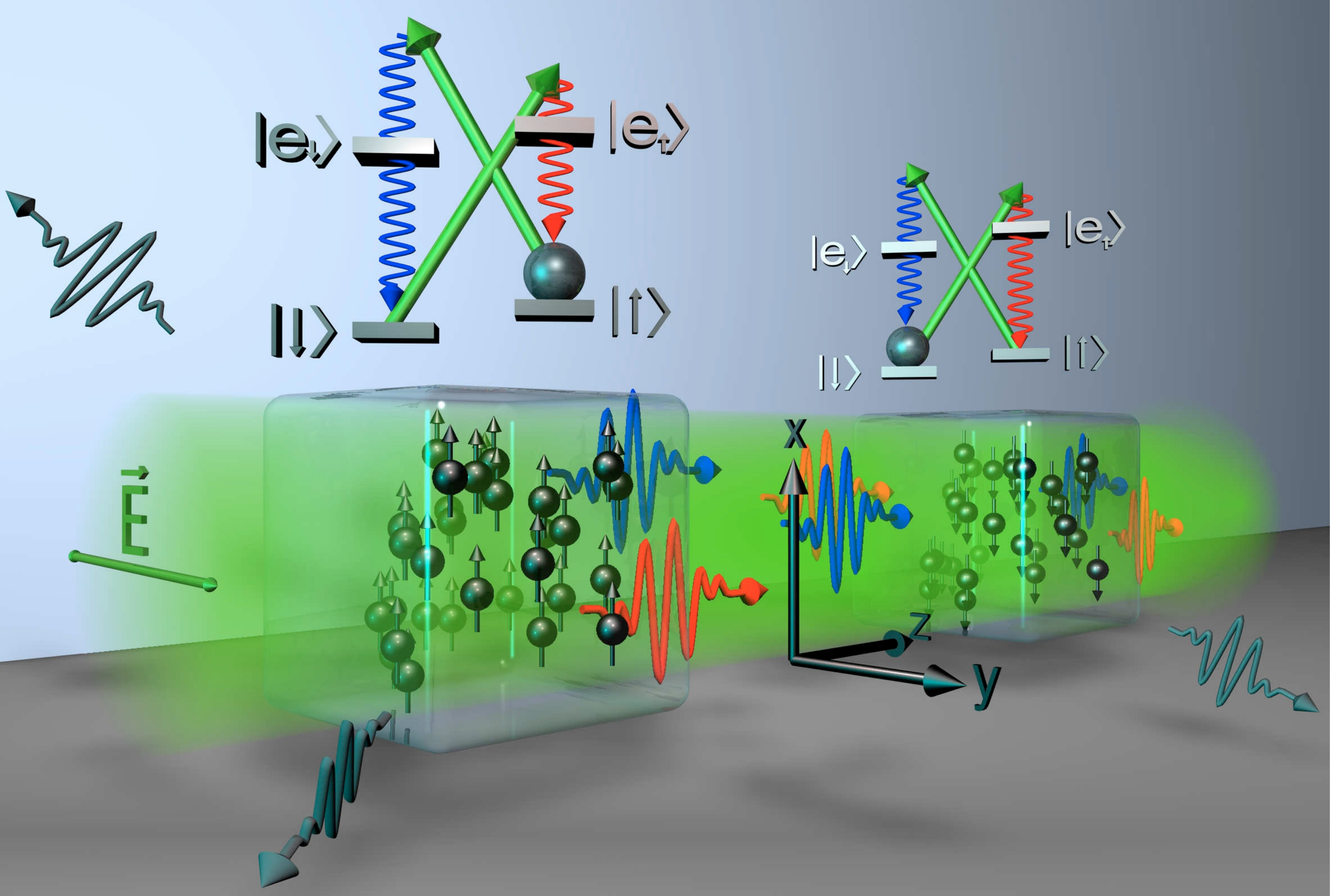}
\caption{Setup for dissipative entanglement generation between two
spatially separated atomic ensembles. A homogeneous magnetic
field, which is applied along the $\hat{x}$-direction defines the
quantization axis and leads to a Zeeman splitting of the atomic
ground states $|\!\!\uparrow \rangle$ and
$|\!\!\downarrow\rangle$. A strong $\hat{y}$-polarized laser beam
(shown in green) couples these states off-resonantly to the
excited states $|e_{\uparrow}\rangle$ and $|e_{\downarrow}\rangle$
and to the light field in $\hat{x}$-direction (shown as wavy
lines). The Zeeman shift $\Omega$ of the ground states leads to
the emission of photons into two sideband modes centered around
$\omega_{\mathrm{L}}\pm\Omega$, where $\omega_{\mathrm{L}}$ is the
frequency of the applied laser field. The vacuum modes in
$\hat{x}$-direction provide a common environment for the two
atomic systems. Due to collective effects, the scattering of
photons in this direction is enhanced.} \label{Fig:Setup}
\end{figure}
In this section, we provide a brief overview to the method of
dissipative entanglement generation put forward in
\cite{EbDExperiment,EbDExperimentArXiv} and \cite{EbDtheory}. We
explain the main idea, introduce the experimental setup and
summarize our main
results.\\
\\ \emph{Key idea}\\
As outlined in the introduction, the entangling mechanism employed
here is invoked by reservoir engineering and drives the system
into a unique inseparable steady state. In the absence of other
decoherence mechanisms, the steady state of the system
(corresponding to the reduced density matrix after tracing out the
environment) can be a pure state. In our case, the reservoir
consists of a continuum of electromagnetic vacuum modes, which
provide a common environment for the two ensembles. The atomic
system can be coupled to this environment in a controlled fashion
by applying suitable laser fields.
More specifically, we consider the setup shown in
Fig.~\ref{Fig:Setup}. Each atomic ensemble consists of a large
number $N$ of hydrogen-like atoms with an internal level structure
with two ground states $|\!\!\uparrow\rangle$ and
$|\!\!\downarrow\rangle$. Both ensembles are driven by a far
off-resonant $\hat{y}$-polarized coherent field. This strong
classical driving field induces effective ground state transitions
$|\!\!\uparrow\rangle\rightarrow
|e_{\downarrow}\rangle\rightarrow|\!\!\downarrow\rangle$ and
$|\!\!\downarrow\rangle\rightarrow
|e_{\uparrow}\rangle\rightarrow|\!\!\uparrow\rangle$, which
involve the emission of $\hat{x}$-polarized photons (compare
Fig.~\ref{Fig:Setup}). This way, the classical driving field
couples the atomic system to the bath of electromagnetic modes in
$\hat{x}$-polarization. The basic entangling mechanism can be
understood by considering the $\hat{x}$-polarized vacuum modes in
the direction of the laser field with wave-vector
$\mathbf{k}_{\mathrm{L}}$ and the rest of the modes separately.
The latter give rise to the standard spontaneous emission and
represent noise processes. The former are shared by both ensembles
and provide therefore the desired common environment. In the
setting considered here, emission into the forward direction is
collectively enhanced for a large optical depth $d$ \cite{HSP10}.
Hence, these modes can successfully compete with all the others
and the entangling processes happen on a faster
time scale than the undesired ones.\\
\\In our case, the target state is a two mode squeezed state
which is entangled in the collective spin states of the two atomic
ensembles $|\Psi_{\mathrm{EPR}}\rangle$. This state is reminiscent
of the entangled quantum state introduced by Einstein, Podolski
and Rosen (EPR) \cite{EPR}. $|\Psi_{\mathrm{EPR}}\rangle$ is the
simultaneous eigenstate with eigenvalue zero of two nonlocal
operators $A$ and $B$,
$A|\Psi_{\mathrm{EPR}}\rangle=0$,
$B|\Psi_{\mathrm{EPR}}\rangle=0$,
where
\begin{eqnarray}\label{Eq:JumpOperators}
A&=&\mu J^-_{\mathrm{I}}-\nu J^-_{\mathrm{II}},\ \ \ \ \ B=\mu
J^+_{\mathrm{II}}- \nu J^+_{\mathrm{I}}.
\end{eqnarray}
$J^{\!\pm}_{\mathrm{I\!/\!II}}$ are collective spin operators with
$J^-\!\!=\!\!\sum_{i}|\!\!\uparrow\rangle_{i}\langle
\downarrow\!\!|$.\\
$\mu=\cosh(r)$ and $\nu=\sinh(r)$, where $r$ is the so-called
squeezing parameter~\cite{FootnoteEPR}. This type of entangled
state is the main working horse for applications in quantum
information science with atomic ensembles and continuous variable
systems in general \cite{BLreview05}. Entanglement can be verified
and quantified by the parameter $\xi=\Sigma_{J}/\left(2|\langle
J_{x}\rangle| \right)=(\mu-\nu)^{2}$, with
$\Sigma_{J}=\mathrm{var} (J_{y,\mathrm{I}}- J_{y,\mathrm{II}})+
\mathrm{var}(J_{\mathrm{z},I} - J_{z,\mathrm{II}})$. $\xi<1$
certifies the creation of an inseparable state
\cite{RFSdG03,JKP01}.\\
The existence of a steady state $|\Psi_{\mathrm{EPR}}\rangle$ with
$A|\Psi_{\mathrm{EPR}}\rangle=B|\Psi_{\mathrm{EPR}}\rangle=0$ can
be understood as an interference process where a
($\hat{x}$-polarized) photon, which is emitted in forward
direction with a given frequency could have been originated from
either of the two ensembles. For a particular atomic state,
$|\Psi_{\mathrm{EPR}}\rangle$, these two process interfere
destructively, such that no photon is scattered into this
direction. As a consequence, the interaction with the laser is
switched off and the state remains the same. This interpretation
can be quantitatively understood in terms of a master equation.
\\In the scheme reported on here, the system-reservoir coupling is
engineered such that it gives rise to a dissipative dynamics which
is governed by
 \begin{eqnarray}\label{Eq:MasterEquation1}
 \frac{d}{dt}\rho\!\!&=&\!\!\cal{L}_{\mathrm{ent}}(\rho)\\
 \!\!&\propto&\!\! \left(\mathrm{A}\rho \mathrm{A}^{\dag} \!-\!\mathrm{A}^{\dag}\mathrm{A}\rho
\!+\!\mathrm{H.C.}\right)\!+\!\left(\mathrm{B}\rho
\mathrm{B}^{\dag} \!-\!\mathrm{B}^{\dag}\mathrm{B}\rho
\!+\!\mathrm{H.C.}\right).\nonumber
 \end{eqnarray}
It can be shown \cite{EbDtheory} that
$\rho_{\mathrm{EPR}}=|\Psi_{\mathrm{EPR}}\rangle\langle
\Psi_{\mathrm{EPR}}|$ is the unique steady state of this
evolution.
In order to obtain a two mode squeezed state, the system is
coupled to two reservoirs, which give rise to the jump operators
$A$ and $B$ respectively. To this end, the ensembles are placed in
a homogeneous magnetic field which causes a Zeeman splitting of
the atomic ground states $\Omega$. Due to the different ground
state energies, photons are scattered into two different frequency
bands centered around $\omega_{\mathrm{L}}\pm \Omega$, which we
will refer to as upper (blue) and lower (red) sideband
respectively. For a sufficiently large separation in frequency
space ($\Omega\gg\Gamma_{\mathrm{Atomic}}$, where
$\Gamma_{\mathrm{Atomic}}$ is the largest effective atomic
transition rate for ground state transitions
$|\uparrow\rangle\leftrightarrow|\downarrow\rangle$
\cite{EbDtheory}), the continua of modes in the lower and upper
sideband can be treated as independent reservoirs
(compare~\ref{App:MasterEquation}). The first (second) term in
Eq.~(\ref{Eq:MasterEquation1}) is due to the interaction with the
photons in the lower (upper) sideband. Note that the jump
operators defined in Eq.~(\ref{Eq:JumpOperators}) are nonlocal
(i.e. involve atomic operators referring to both, the first and
the second ensemble) and can give therefore rise to an entangled
steady state. As explained above, it originates from the
interference between processes where a photon is emitted in
forward direction by the first or the second ensemble.
Other processes can be included in the form of additional terms in
the master equation
$ \frac{d}{dt}\rho=\cal {L}_{\rm ent}(\rho) + {\cal L}_{\rm
noise}(\rho)$,
where ${\cal L}_{\rm noise}(\rho)$, summarizes undesired processes
such as spontaneous emission, collisions and fluctuating magnetic
fields. A key point lies in the fact that the rate of the
entangling processes ($\cal {L}_{\rm ent}(\rho)$) scales with the
optical thickness (due to collective effects which originate from
constructive interference involving each individual atom within a
single ensemble), whereas the rate of the detrimental processes
does not. Thus, for sufficiently optically thick samples, the
creation of entanglement in a steady state is possible even in the
presence of noise (see Eq.~(\ref{Eq:SteadyState}) in
\ref{App:CalculationEntanglement}).\\
\\ \emph{Setup and experimental results}\\
The experiment is carried out using $^{133}$Cs vapor at room
temperature. The two-level subsystem is encoded in the two
outermost hyperfine levels of the $6S_{1/2}$ ground state within
the manifold with total spin $F=4$~\cite{FootnoteCesium}. We
identify $|\!\!\uparrow\rangle_I\equiv|F=4,m_F=4\rangle$,
$|\!\!\downarrow\rangle_I\equiv|F=4,m_F=3\rangle$ and
$|\!\!\uparrow\rangle_{II}\equiv|F=4,m_F=-3\rangle$,
$|\!\!\downarrow\rangle_I\equiv|F=4,m_F=-4\rangle$ (where $m_F$ is
the magnetic quantum number).
The atoms are confined in cubic glass cells which are separated by
a distance of approximately $0.5$m and have a spatial extent of
$2.2$cm. Each cell contains $10^{12}$ atoms and is equipped with a
paraffin-based spin-preserving coating. The experimental setup is
sketched in Fig. 2a. The two ensembles are prepared in oppositely
oriented coherent spin states (CSS). This is achieved by optically
pumping the atoms of the ensembles into $m_F =\pm4$ in the $\hat
x$-direction respectively. The circularly polarized pump lasers
are depicted in blue and Fig. 2b shows the atomic level structure,
indicating laser frequencies and polarization. The strong probe
beam which is initially polarized in $\hat y$-direction
transverses the atoms in the $\hat z$-direction. Behind the cells,
the detection system is set up.
\par
\begin{figure}
\centering
\includegraphics[width=\columnwidth]{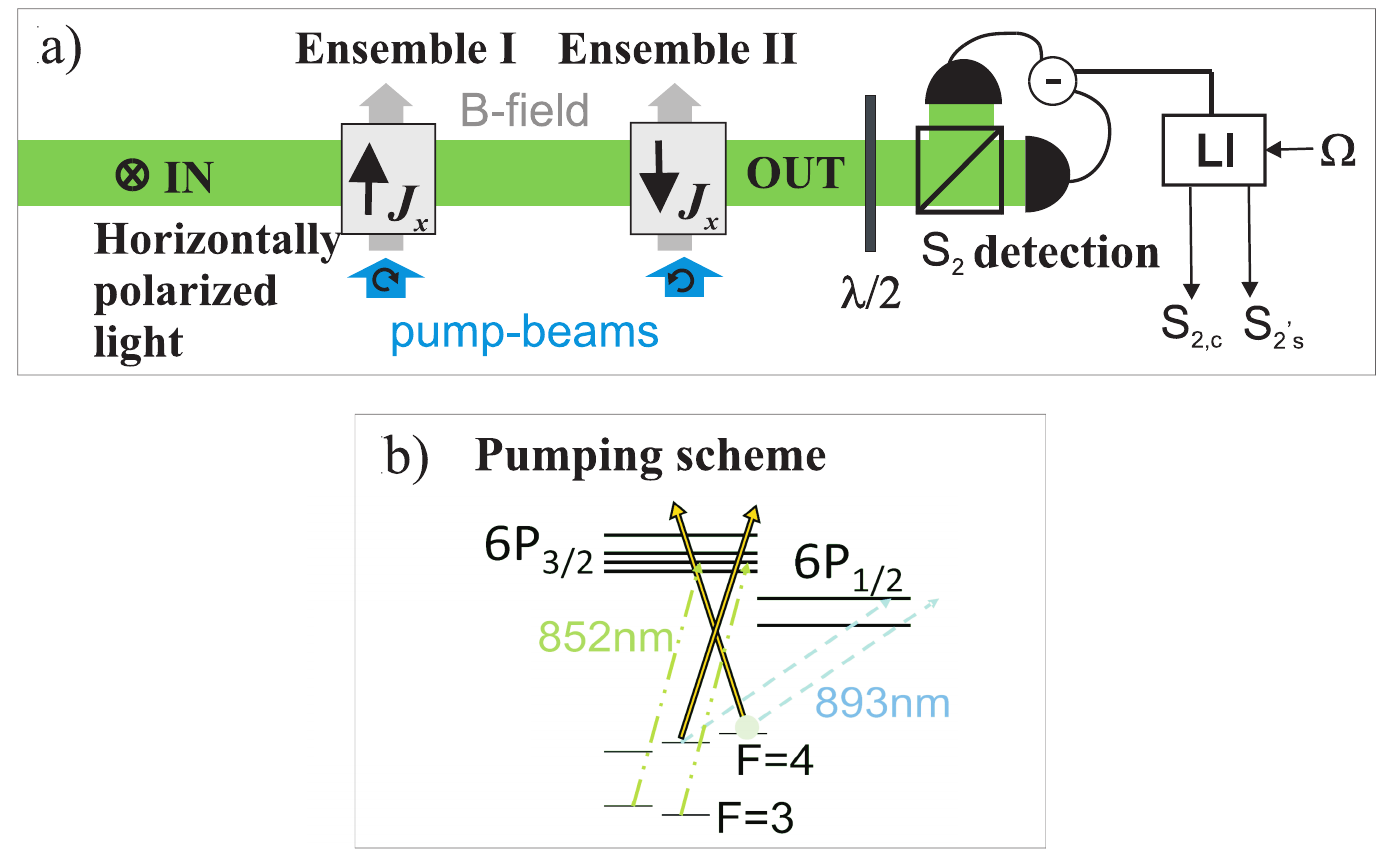}
\caption{a) Experimental setup. The horizontally polarized probe
light shown in green passes through two oppositely oriented atomic
ensembles. The orientation is generated via optical pumping. The
optical pumping beams are depicted in blue. Behind the two atomic
cells, the $S_2$ detector signal is processed by the lock-in
amplifier (LA) to take measurements on the atomic quantum spin
components $J_{y,z}$ in the rotating frame. b) shows the optical
pumping scheme. The relevant laser frequencies and polarizations
are indicated in a level scheme.}\label{Fig:Exp}
\end{figure}
In \cite{EbDExperiment}, two types of results have been obtained.
Firstly, entanglement has been created purely dissipatively,
demonstrating that this type of processes can be harnessed for
tasks in Quantum Information Science. In this series of
experiments, entanglement has been maintained for a time span,
which is an order of magnitude longer than the time intervals for
which entanglement could be sustained in this system so far
\cite{HSP10}.
Entanglement has been obtained in a quasi steady state rather than
in a true steady state due to the multi-level structure of Cesium.
As opposed to the two-level model discussed above, atoms can leave
the two-level system by undergoing transitions to other internal
states. The entangling processes are fast compared to these
undesired transitions. The desired dynamics with respect to the
two-level subsystem reaches a steady state, but since it is
superposed by the slow detrimental processes involving atom
losses, entanglement disappears.\\
In a second series of experiments, strong pump and repump fields
have been applied in order to transfer the atoms back, which left
the two-level system. These incoherent fields lead to increased
noise contributions, which prevents the generation of a purely
dissipative steady state in this particular setting. However, by
combining the dissipative mechanism with continuous measurements
on the light field, a true steady state has been obtained.
%
%
In the following, we consider the generation of entanglement by
dissipation in more detail.

\section{Purely dissipative
entanglement}\label{Subsec:PurelyDissipativeEntanglement}
This section is concerned with the purely dissipative generation
of entanglement. The basic mechanism is explained and further
substantiated by additional experimental data. Moreover, the
perspectives for obtaining steady state entanglement for
multi-level systems is discussed.\\
\\ \emph{Entanglement creation by virtue of interference}\\
As mentioned earlier, the underlying mechanism can be understood
as an interference effect in the dissipative channel. Due to
destructive interference, no photon is scattered into the forward
direction in the steady state. Accordingly, no measurement needs
to be performed on the light field.
\begin{figure}
\centering
\includegraphics[width=6cm]{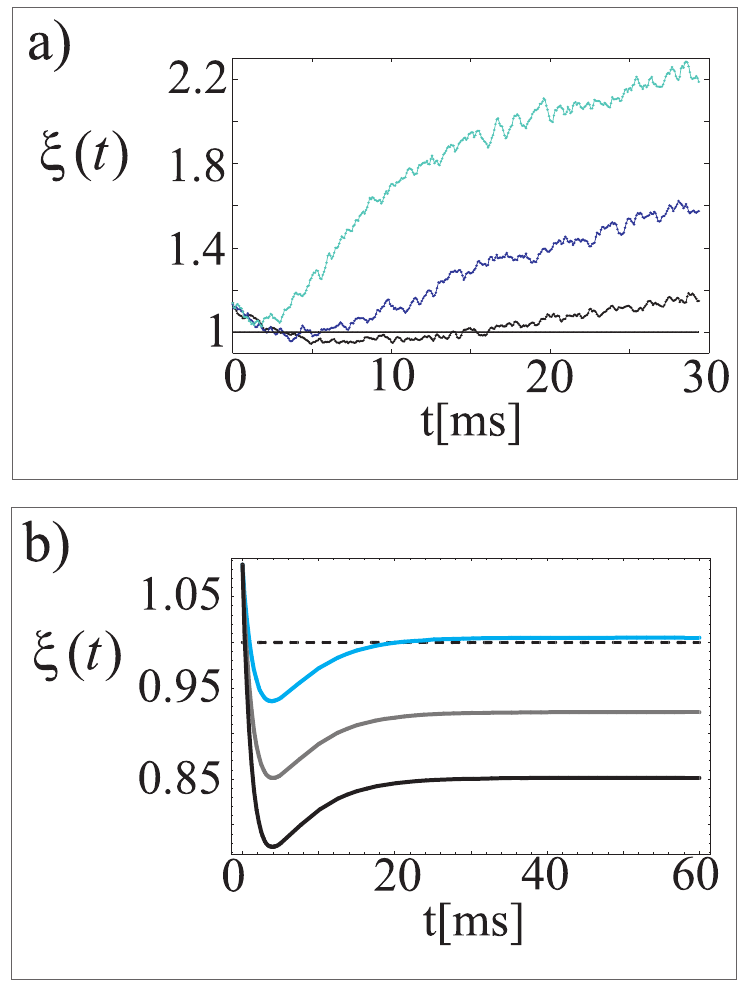}
\caption{Entanglement generated by dissipation a) Entanglement for
different detunings $\delta \Omega$ of the magnetic fields. The
different curves correspond to $\delta \Omega=0$ (black), $\delta
\Omega=20$Hz (violet) and $\delta \Omega=40$Hz (blue). b)
Dissipatively generated steady state entanglement. Predicted time
evolution of $\xi(t)$ in the presence of strong pump and repump
fields for $d=55$, (blue), $d=100$ (grey) and $d=150$ (black). The
other parameters take values close to the ones used in
\cite{EbDExperiment}, as explained in \ref{App:FitData}.}
\label{Fig:PureDissipation}
\end{figure}
Additional evidence that the entanglement is produced by the
collective dissipation process is provided by its dependence on
the dephasing between spin coherences of the two ensembles. To
demonstrate this dependence, we introduce a detuning $\delta
\Omega=\Omega_{I}-\Omega_{II}$ of the  Larmor frequencies of the
two ensembles by tuning the bias magnetic fields.
Fig.~\ref{Fig:PureDissipation}a shows that already at $\delta
\Omega =20$Hz entanglement disappears. This can be understood as a
consequence of the "which way" information (or "which ensemble"
information in this case) provided by the distinguishability of
the photons scattered from the two ensembles into the sideband
modes $\omega_{\mathrm{L}}\pm \Omega_{\mathrm{I}}$ and
$\omega_{\mathrm{L}}\pm \Omega_{\mathrm{II}}$. Hence the atomic
samples do not share the same reservoir any more and entanglement
disappears.\\
Note that the effect demonstrated in the experiment can not be
understood in terms of photons which are spontaneously emitted by
the first ensemble interacting with the second one. Due to the
large detuning ($\Delta=850$MHz), the interaction of a single
photons emitted by the first ensemble with atoms in the second one
is negligible in the parameter regime accessible in the
experiment. Here, atomic transitions are to a very good
approximation only induced by the classical driving field
(accordingly, the input-output relations for atoms in the second
sample are identical with those in the absence of the first
ensemble).\\
\\ \emph{Comparison of dissipative entanglement generation with other methods}\\
There exists a large variety of methods for creating entanglement
between two quantum systems. In particular, several methods have
been devised and demonstrated for entangling atomic ensembles.
Even though some schemes, which have been experimentally
implemented share similarities with the one described here, they
are fundamentally different. In the following, we explain this in
detail and highlight the features of the method realized here in
comparison with previously demonstrated ones.\\
In standard approaches, which are based on a coherent interaction
followed by a measurement
\cite{DuCZP00,DLCZ01,JKP01,ChMJYKK05,ChRFPEK05,EiAMFZL05,MCJ06,CDLK08,YCZC08,CCYZ08,K08,HSP10},
two atomic ensembles A, and B are prepared in specific pure states
$|a\rangle_{\mathrm{A}}$, and $|b\rangle_{\mathrm{B}}$. An
additional system, E, which typically corresponds to certain modes
of the electromagnetic field, is also initialized in a specific
state, for example the vacuum $|0\rangle_{\mathrm{E}}$. For
appropriately chosen external parameters such as the frequency and
polarization of applied laser fields, the interaction of system E
with A and B gives rise to an entangled state $|\Psi\rangle= U
|a,b\rangle_{\mathrm{A,B}}|0\rangle_E$. If system E is measured,
e.g. using a beam splitter and single-photon detectors, or by
means of homodyne detection, the state of systems A and B is
projected onto an entangled state,
$|\Phi(e)\rangle_{\mathrm{A,B}}$. This state depends on the
outcome of the measurement, $e$. If no measurement is performed
(which corresponds to averaging with respect to the possible
measurement outcomes \cite{FootnoteMeasurement}) the resulting
state is not entangled. For instance, the DLCZ protocol
\cite{DLCZ01}, yields a separable state if the photons emitted by
the ensembles are not detected.\\
\\Dissipative methods can be described as follows.
$|a\rangle_\mathrm{A}$, $|b\rangle_\mathrm{B}$, and
$|0\rangle_\mathrm{E}$ denote again the initial states of systems
A, B and E respectively. Due to the interaction of system E with A
and B, the state $|\Psi(t)\rangle=
U(t)|a,b\rangle_{A,B}|0\rangle_E$ is created, where the dependence
of the resulting quantum state on the time $t$ is explicitly
indicated. Under ideal conditions, i.e. if systems $A$ and $B$ do
not couple to other environments, the interaction of A and B with
E can be engineered such that the atomic system evolves towards an
entangled state. In contrast to the schemes described above, the
implementation of this entangling dynamics does not require
measurements on system E. This type of behavior can occur if
system E possesses an infinite number of degrees of freedom, such
that a non-unitary dynamics drives the system towards a fixed
state. Due to this property, E is typically referred to as
environment and the corresponding interaction with systems A and B
is referred to as dissipative process. Dissipative phenomena of
this kind are best described by means of master equations. To this
end, the environment is traced out and an equation for the reduced
density operator of systems A and B, $\rho$, is derived as
described in \ref{App:MasterEquation}.
In the presence of other environments, the dissipation induced by
the coupling of A and B to system E can still create entanglement
with a life time, which exceeds the decoherence times due to these
extra noise sources significantly, if the corresponding
(uncontrolled) coupling is sufficiently weak. Note further, that
typically, noise processes can be included in the master equation
description as it is done in the present work.\\
\\In the experiment discussed here, entanglement induced by
dissipation has been observed. In particular, in contrast to
approaches which have been previously implemented, entanglement is
obtained without using measurements on the quantum state of the
environment~\cite{FootnoteRemark}. Furthermore, systems A and B
remain entangled for $40$ms. This entanglement life-time is at
least by a factor $16$ longer than the decoherence time induced by
other noise sources. It has been experimentally verified that in
the absence of the dissipative process, the measured entanglement
life time is limited to $2.5$ms due to the remaining noise sources
such
as collisions or inhomogeneities of the applied magnetic fields.\\
Dissipative methods exhibit another distinctive feature, which is
present for an ideal two--level system (compare \cite{EbDtheory}),
but not in the multilevel description of the experiment. For long
times $t \rightarrow \infty$, systems A and B decouple from the
environment E, $|\Psi(t)\rangle\to |\Phi\rangle_{A,B}
|E(a,b,t)\rangle_E$ under ideal conditions, i.e. in the absence of
additional noise sources. Remarkably, the desired state
$|\Phi\rangle_{A,B}$  is reached irrespective of the initial state
of systems A and B which can be highly mixed. Moreover, except for
an initial waiting time, no special timing is required. This
behavior is again due to the fact that E possesses an infinite
number of degrees of freedom, which guarantees that revival
effects are not present. This way, entropy is transferred from the
system to the environment, which drives A and B into a particular
steady state, which depends only
on the engineered coupling.\\
Using dissipative methods, a mixed but still entangled steady
state can be reached even in the presence of additional noise
sources, as long as the coupling of A and B to other environments
is sufficiently weak compared to the engineered dissipative
processes. This opens up the possibility to keep systems A and B
entangled for arbitrarily long times.\\
\\ \emph{Perspectives for creating purely dissipative steady state
entanglement in multi-level systems.}\\
Below, we investigate the possibility of generating purely
dissipative steady state entanglement in atoms with multi-level
ground states. The main reason, why the system in the experiment
reported on in \cite{EbDExperiment} does not display purely
dissipatively generated entanglement in a steady state is the
depopulation of the relevant two-level subsystem due to
spontaneous emission which transfers the atoms into other Zeeman
levels. The depopulation of the relevant levels can be avoided by
applying strong laser fields, which transfer atoms back. However,
these fields introduce additional decoherence processes which
inhibit the creation of entanglement in a steady state. This
problem can be circumvented by increasing the optical depth of the
atomic ensembles such that the entangling dissipative process
prevails over the noise processes and dominates the dynamics.\\
We consider this scenario by including additional $\sigma_{\pm}$
polarized pump and repump fields, which induce resonant
transitions with $\Delta m_{F}\pm 1$ in the first/second ensemble
in the model. Pump fields drive transitions within the manifold of
atomic states with $F=4$ and repump fields transfer atoms from
states with $F=3$ back to $F=4$. As explained in
~\ref{App:FitData}, we estimate the effect of these fields using a
simplified model, which has been used in~\cite{EbDExperiment} to
fit the experimental data.
Fig.~\ref{Fig:PureDissipation}b shows the predicted time evolution
of entanglement for $d=55;100;150$ in the presence of both, pump
and repump fields (in \cite{EbDExperiment}, an optical depth of
$d=55$ was used). The other parameters used in this calculation
take values close to the ones used in \cite{EbDExperiment}. The
theory predicts that under the present maximal optical depth
$d=55$, the steady state atomic variance is just slightly above
the separability criterion (this has also been confirmed
experimentally). However, the experimental realization of purely
dissipative steady state entanglement should be feasible along two
possible routes. Firstly, atoms possessing two-level electronic
ground states can be used, for example Ytterbium
($^{171}\mathrm{Yb}$)~\cite{TTI06,TFNT09} . In this case, the
two-level theory formulated here can be directly implemented,
avoiding additional dynamics which leads to the growth of $\xi$
with time. Alternatively, a true dissipatively generated steady
state using multi-level ground states can be achieved for higher
optical depths, which can be obtained, for example, by placing the
atoms inside a low finesse optical cavity.\\
In \cite{EbDExperiment}, we devised and implemented an alternative
approach, which enables the creation of entanglement which
persists for arbitrarily long times. This alternative approach
combines the dissipative mechanism with continuous measurements as
explained below.

\section{Dissipative entanglement assisted by
measurements}\label{Sec:EntanglementAssistedByMeasurements}
\begin{figure}
\includegraphics[width=6cm]{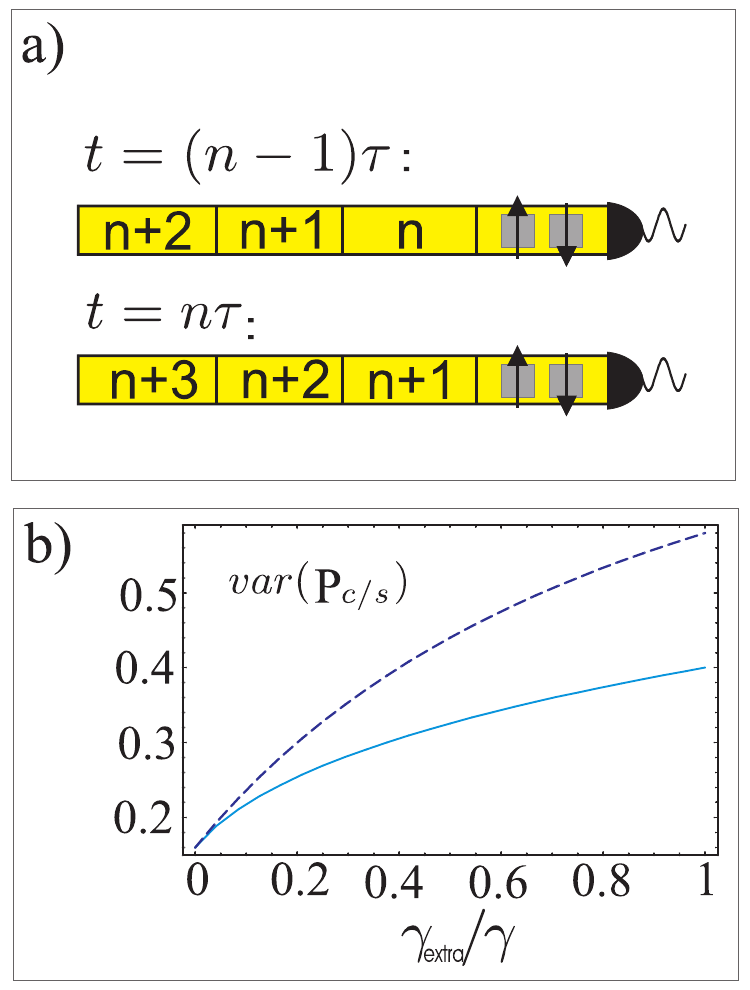}
\caption{\textbf{Steady state entanglement assisted by
measurements.} a) Illustration of the interaction of atoms and
light in terms of spatially localized modes b) Squeezed atomic
variance $\mathrm{var}\left(P_{c/s}\right)$ in the steady state
versus the ratio $\gamma_{\mathrm{extra}}/\gamma_s$ in the absence
of measurements (dashed line) and if the $y$-quadrature of the
scattered light field is measured (full line).
$\gamma_{\mathrm{extra}}$ and $\gamma_s$ denote the rates of the
desired entangling processes and the atomic decay respectively.}
\label{Fig:TheoryEbDwithMeasurements}
\end{figure}

In the following, we show by means of a simple model how
measurements on the light field can improve the generation of
entanglement under the dissipative dynamics described above in the
presence of noise sources. In the basic model employed here to
illustrate the relevant effects, the Holstein-Primakoff
approximation \cite{HoP40} is used to describe the atomic system
and noise is included in the form of decay of the transverse spin
components $J_{y}$ and $ J_{z}$ at a rate
$\gamma_{\mathrm{extra}}$ (a more detailed discussion is to be
published elsewhere). In this subsection, the dissipative
generation of entanglement assisted by measurements is explained
in terms of input-output relations, since this approach is more
illustrative than the master equation formalism employed in
Sec.~\ref{Sec:Overview}. Both descriptions are equivalent and
yield the same
results.\\
\\For large, strongly polarized atomic ensembles, collective spins
can be described by bosonic modes in terms of the quadratures
\begin{eqnarray*}
X_{I/II}&=&J_{y,I/II}/\sqrt{|\langle J_{x,I/II}\rangle |},\\
P_{I/II}&=&\pm J_{z,I/II}/\sqrt{|\langle J_{x,I/II}\rangle|}.
\end{eqnarray*}
Light propagates in $\hat{z}$-direction (see Fig.~\ref{Fig:Setup})
and interacts with the atomic ensembles. We consider here a
one-dimensional model which includes only light scattered in
forward direction. Processes corresponding to scattering of
photons into other directions enter in the form of noise.
Light is characterized in terms of spatially localized modes
\cite{SiD03,MaM04}
\begin{eqnarray*}
y(z)&=&\frac{1}{\sqrt{4\pi}}\int_b d\omega
\left(a_{\omega}e^{ \frac{i}{c}\left(\omega-\omega_{\mathrm{L}}\right) z}+H.C.\right),\\
q(z)&=&\frac{-i}{\sqrt{4\pi}}\int_b d\omega
\left(a_{\omega}e^{ \frac{i}{c}\left(\omega-\omega_{\mathrm{L}}\right) z}-H.C.\right),\\
\end{eqnarray*}
where $c$ is the speed of light. $b$ and $\omega_{\mathrm{L}}$ are
the bandwidth and central frequency of the applied laser field.
The operators for spatially localized field modes $y(z)$ and
$q(z)$ obey the canonical commutation relation
$[y(z),q(z')]=ic\delta_b(z-z')$, where the deltafunction has a
width of the order $c/b$. The spatial argument $z$ refers to the
distance along the propagation direction $\hat{z}$.
Atoms and light interact according to
$ H=H_{\mathrm{A}}+H_{\mathrm{L}}+H_{\mathrm{int}}, $
where
$H_{\mathrm{A}}=\frac{\Omega}{2}\left(X_{I}^2+P_{I}^2\right)-\frac{\Omega}{2}\left(X_{II}^2+P_{II}^2\right)$
describes the Zeeman splitting of the atomic ground states. Due to
the applied magnetic field, the transverse components of the
collective spin described by $X_{\mathrm{I/II}}$ and
$P_{\mathrm{I/II}}$ precess at the Larmor frequency $\Omega$.
$H_{\mathrm{L}}$ represents the free propagation of light
$\frac{d}{dt}y(z)=i[H_{\mathrm{L}},y(z)]\cong -c
\frac{d}{dt}y(z)$. The interaction of the light field with two
pointlike atomic ensembles \cite{HaPC05} located at $z=0$ and
$z=R$ is given by
\begin{eqnarray*}
H_{\mathrm{int}}&=&\sqrt{2\gamma_s}\left(\frac{1}{Z}
X_{I}y(0)+ZP_{I}q(0)\right)\\
&+&\sqrt{2\gamma_s}\left(\frac{1}{Z}
X_{II}y(R)+ZP_{II}q(R)\right),
\end{eqnarray*}
where $Z=\mu+\nu$ and $1/Z=\mu-\nu$. We assume in the following,
that the distance $R$ between the ensembles can be neglected,
which is justified for $\Gamma_{\mathrm{Atomic}}R\ll c$, where
$\Gamma_{\mathrm{Atomic}}$ is the effective rate at which
transitions between ground states occur, and $L^2 k_L \gg R$,
where $L$ is the spatial extend of an atomic ensemble and $k_L$ is
the wave vector of the driving field. The former is a necessary
condition to neglect retardation effects, while the latter is used
in the calculation of averaged emission rates for fast moving
atoms (compare \ref{App:MasterEquation} and \cite{EbDtheory}).
Both conditions are well fulfilled for the parameter regime of the
experiment, where the effective decay rates are of the order of
few ms and $R$ is on the order of a meter ($L$ is about $2$cm and
$k_L$ on the order of $10^7$m$^{-1}$).
The resulting Heisenberg equations can be solved by changing to a
coordinate system rotating at the Larmor frequency $\Omega$ and
performing the variable transformation $\zeta=ct-z$ such that
$\bar{y}(\zeta,t)=y(ct-\zeta,t)$. We introduce exponential
$\cos(\Omega t)$ and $\sin(\Omega t)$ modulated light functions
\begin{eqnarray}
y_{c,\pm}^{\mathrm{in}}&=&\frac{1}{\mathcal{N}_\pm}\int_{0}^{T}e^{\pm
\gamma_s
t}\cos(\Omega t)\bar{y}(\zeta,0),\nonumber\\
q_{c,\pm}^{\mathrm{in}}&=&\frac{1}{\mathcal{N}_{\pm}}\int_{0}^{T}e^{\pm
\gamma_s t}\cos(\Omega t)\bar{q}(\zeta,0),\label{yq-def}
\end{eqnarray}
with $\mathcal{N}_+=\frac{2\sqrt{\gamma_s}}{\sqrt{e^{2\gamma_s
T}-1}}$ and
$\mathcal{N}_-=\frac{2\sqrt{\gamma_s}}{\sqrt{1-e^{-2\gamma_s
T}}}$. $y_{s,\pm}^{\mathrm{in}}$ and $q_{s,\pm}^{\mathrm{in}}$ are
defined analogously. We assume that the Larmor precession is fast
compared to the atomic evolution ($\Omega T\gg 1$), such that the
operators describing the sin and cos modulated light modes are
canonical and independent,
$[y_{c,\pm},q_{c,\pm}]=[y_{s,\pm},q_{s,\pm}]=i$ and
$[y_{c,\pm},q_{s,\pm}]=[y_{s,\pm},q_{c,\pm}]=0$. This yields
\begin{eqnarray}
X_{c/s}^{\mathrm{out}}&=&e^{-\gamma_{s}T}X_{c/s}^{\mathrm{in}}+Z\sqrt{1-e^{-2\gamma_{s}T}}q_{c/s,+}^{\mathrm{in}},\nonumber\\
P_{c/s}^{\mathrm{out}}&=&e^{-\gamma_{s}T}P_{c/s}^{\mathrm{in}}-\frac{1}{Z}\sqrt{1-e^{-2\gamma_{s}T}}y_{c/s,+}^{\mathrm{in}},\nonumber\\
y_{c/s,-}^{\mathrm{out}}&=&e^{-\gamma_{s}T}y_{c/s,+}^{\mathrm{in}}+Z\sqrt{1-e^{-2\gamma_{s}T}}P_{c/s}^{\mathrm{in}},\nonumber\\
q_{c/s,-}^{\mathrm{out}}&=&e^{-\gamma_{s}T}q_{c/s,+}^{\mathrm{in}}-\frac{1}{Z}\sqrt{1-e^{-2\gamma_{s}T}}X_{c/s}^{\mathrm{in}},\label{io-eq}
\end{eqnarray}
where $X_{c}=\frac{1}{\sqrt{2}}\left(X_{I}+X_{II}\right)$,
$P_{c}=\frac{1}{\sqrt{2}}\left(P_{I}+P_{II}\right)$ and
$X_{s}=\frac{-1}{\sqrt{2}}\left(P_{I}-P_{II}\right)$,
$P_{s}=\frac{1}{\sqrt{2}}\left(X_{I}-X_{II}\right)$ was used.
As a next step, we include continuous measurements on the light
field and consider the corresponding time evolution of the atomic
state in the Schr\"odinger picture. The continuous interaction and
measurement process shown in Fig.~\ref{Fig:Setup} is illustrated
schematically in Fig.~\ref{Fig:TheoryEbDwithMeasurements}a in a
discretized way. Spatially localized light modes correspond here
to infinitesimally short pulses of duration $\tau\sim 1/b$ (where
$b$ is the bandwidth of the incident laser field as explained
above), which interact successively with the atomic system. Each
of these spatially localized light modes is initially in the
vacuum state, such that the quantum state at time $t=n \tau$ is
given by $|\Psi(t)\rangle_{A}|0\rangle_{L,n+1}$, where
$|\Psi(t)\rangle_{A}$ denotes the atomic state at time $t$. Then
atoms and light are subject to an entangling interaction resulting
in the quantum state $e^{-i H
\tau}|\Psi(t)\rangle_{A}|0\rangle_{L,n+1}$. Finally, the
$y$-quadrature of the light field is measured, yielding the
measurement outcome $y_n$ such that
$|\Psi(t+\tau)\rangle_{A}=\frac{1}{\sqrt{P(y_n)}}\
_{\mathrm{L,n+1}}\!\langle y_n|e^{-i H
\tau}|0\rangle_{L,n+1}|\Psi(t)\rangle_{A}$, where $P(y_n)$ is the
probability to obtain the result $y_n$. The resulting expression
can be expanded up to first order in the parameter $\tau$ yielding
a differential equation for the time evolution of the atomic
system. The atomic state obtained after the measurement depends on
the measurement outcome $y_n$.
We consider here Gaussian quantum states, i.e. states with a
Gaussian Wigner function. These states are fully characterized by
their first and second moments, - prominent examples include
coherent as well as two mode squeezed states. The Gaussian
character of a state is preserved under the evolution of
Hamiltonians, which are at most quadratic in the system operators
and under Gaussian measurements such as homodyne detection. Since
we consider all interactions and measurements to be of this kind
and all states to be Gaussian, the entanglement of the resulting
state is completely determined by the atomic variance
$\mathrm{var}\left(P_{c/s}\right)$, which does not depend on $y_n$
\cite{GC02}. Therefore the resulting entanglement is independent
of the measurement outcome. If the measurement results are traced
out ($\rho(t+\tau)=\sum_{y_n}M_n \rho(t)M_n^{\dag}$, where
$M_n=_{L,n+1}\!\langle y_n|e^{-i H \tau}|0\rangle_{L,n+1}$), and
the resulting expression is evaluated to first order in $\tau$,
the master equation used in Sec.~\ref{Sec:Overview} and
\ref{App:MasterEquation} is recovered, if the spin operators are
replaced by creation and annihilation operators within the
Holstein-Primakoff approximation \cite{HoP40}.
The whole process can be conveniently described by means of the
Gaussian formalism, where atomic states are expressed in terms of
their covariance matrix $\Gamma_c$, and displacement vector
$\mathbf{D}$, which display the second moments (variances and
covariances) and first moments (mean values) of the system
respectively.
In particular, this formalism allows one to easily calculate the
variances of atomic quadratures after the Gaussian measurement of
the $y$-quadrature of the light field,
$\mathrm{var}\left(P_{c/s}\right)_{\mathrm{{cond}}}$ at the end of
each time step depending on the variance prior to the measurement
\begin{eqnarray}
\mathrm{var}\left(P_{c/s}\right)_{\mathrm{{cond}}}=\mathrm{var}(P_{c/s})-\frac{\langle
P_{c/s} y_{c/s} + y_{c/s}
P_{c/s}\rangle^2}{4\mathrm{var}(y_{c/s})},\label{ConditionalVariance}
\end{eqnarray}
where $y_{c/s}$ and $q_{c/s}$ refer here to the localized light
mode interacting with the ensemble in the n$^{\mathrm{th}}$ time
step and $\gamma_s\tau\ll 1$ is assumed. This way, a differential
equation for the squeezed atomic variances is derived. In the
ideal case,
\begin{eqnarray*}
\mathrm{var}\!\left(P_{c/s}\right)_{\mathrm{{cond}}}(t\!+\!\tau)\!\!&=&\!\!\mathrm{var}\!\left(P_{c/s}\right)_{\mathrm{{cond}}}\!(t)
\!+\!\mathrm{var}\!\left(P_{c/s}\right)_{\mathrm{{cond}}}\!(t)\\
\!\!&&\!\!\left(1-\mathrm{var}\left(P_{c/s}\right)(t)Z^2\right)\gamma_s\tau,
\end{eqnarray*}
which yields
\begin{eqnarray*}
\mathrm{var}\!\left(P_{c/s}\right)_{\mathrm{{cond}}}\!(t)\!\!&=&\!\!\frac{1}{e^{-\gamma_{s}t}\!
\left(\mathrm{var}\!\!\left(P_{c/s}\right)\!(0)\right)^{-1}\!\!+\!Z^2\left(1
\!-\!e^{-\gamma_{s}t}\right)},
\end{eqnarray*}
whereas in the absence of measurements,
\begin{eqnarray*}
\mathrm{var}\left(P_{c/s}\right)(t)&=&e^{-\gamma_{s}t}\
\mathrm{var}\left(P_{c/s}\right)\!(0)+\frac{1}{Z^2}\left(1-e^{-\gamma_{s}t}\right).
\end{eqnarray*}
Both time evolutions result in a steady state with
$\mathrm{var}\left(P_{c/s}\right)_{\infty}=1/Z^2=(\mu-\nu)^{2}$,
since atoms and light decouple for $t\rightarrow\infty$.
Accordingly, the steady state entanglement can not be improved by
means of measurements on the light field in the ideal case.
The situation is different in the presence of noise sources, which
prevent the decoupling of atoms and light. In this case, residual
atom-light correlations persist in the steady state and
measurements on the light field can be used to improve the
entanglement. Here, we illustrate this effect by including atomic
transverse decay at a rate
$\gamma_{\mathrm{extra}}$~\cite{FootnoteNoise}.
If the $y$-quadrature of the scattered light field is measured,
one obtains
\begin{eqnarray*}
\xi_{\text{\tiny{{cond}}},{\infty}}\!\!\!&=&\!\!\!\frac{1}{2Z^2}\!\!\left(\!1\!-\!\frac{\gamma_{\text{\tiny{extra}}}}{\gamma_{s}}\!+\!\sqrt{\left(1\!-\!\frac{\gamma_{\text{\tiny{extra}}}}{\gamma_{s}}\right)^2\!+\!4Z^2\frac{\gamma_{\text{\tiny{extra}}}}{\gamma_{s}}}
\right).
\end{eqnarray*}
Fig.~\ref{Fig:TheoryEbDwithMeasurements}b shows that for
$\gamma_{\mathrm{extra}}>0$, the steady state entanglement
described by this equation is higher than the steady state if no
measurements are performed, which is given by
\begin{eqnarray*}
\xi_{\infty}&=&\frac{\frac{1}{Z^2}\gamma_{s}+\gamma_{\mathrm{extra}}}{\gamma_{s}+\gamma_{\mathrm{extra}}}.
\end{eqnarray*}
Note that in principle, all measurement results $y(t)$ obtained
during the continuous measurement procedure could be used to
perform feedback operations which stabilize the atomic state at a
certain position in phase space. However, this is not necessary
here, since the atomic quantum state at time $t$ depends only on
the recent history of the measurements, i.e. on $y(t')$ for
$t_{ss} \le t'\le t$, where $t_{ss}$ is the time it takes to reach
the steady state. For the dissipative processes considered here,
the atomic state $\rho(t)$ is memoryless regarding events which
occurred in a time interval longer than $t_{ss}$. Thus, only
measurement results obtained during a fixed time interval
$t_{ss}$, which is independent of $t$ are needed to localize the
atomic state in phase space.
\section{Experimental atomic state reconstruction}\label{Sec:ExperimentalDetails}
The experimental setup is depicted in Fig.~\ref{Fig:Exp} and
described in the caption and the surrounding text. This section is
focussed on the characterization of the atomic state via
measurements on light which has interacted with the atomic
ensembles.\par The light observable of interest is the Stokes
operator $S_2$ given by the difference of the number of photons
polarized in $\pm 45^{\circ}$. In the specific setting discussed
here, the probe beam is strongly polarized in $\hat y$-direction
and  the $\hat x$-polarization represents the polarization mode of
interest for measurement as discussed in
Sec.~\ref{Sec:EntanglementAssistedByMeasurements}. Then the Stokes
operator $S_2\approx\sqrt{\phi/2}\cdot y$, where $\phi$ is the
photon flux. $S_2$ can be measured with polarization homodyning
techniques as depicted in Fig.~\ref{Fig:Exp}. The light beam is
send through a halfwave plate and a polarizing beam splitter
(PBS). The signals from the detectors situated in the output ports
of the PBS are subtracted and the difference signal is analyzed at
the Larmor frequency $\Omega$ with a lock-in amplifier, since we
are interested in detecting a signal from the spins in the
rotating frame. Additionally the measurement outcome can be
weighted with suitable mode functions $f(t)$ and after suitable
normalization we are thus able to measure the light observables
$y_{c,s-}$ or $y_{c,s+}$  defined in Eq.~(\ref{yq-def}).

\subsection{Atomic state reconstruction}
Now, the variances of the collective atomic operators
$P_c^\mathrm{in}=1/\sqrt 2(P_I^\mathrm{in}+P_{II}^\mathrm{in})$
and $P_s^\mathrm{in}=1/\sqrt
2(X_I^\mathrm{in}-X_{II}^\mathrm{in})$ can be  found from
measurements on the transmitted light. From the variances of the
outgoing operators $y_{c,s-}$, using the input-output relations
given in Eq.~(\ref{io-eq}), it follows:
\begin{eqnarray}
\var(P_c^\mathrm{in})=\frac{1}{\kappa^2}(\var(y_{c-}^\mathrm{out})-\sigma_{in}^2(1-\frac{\kappa^2}{Z^2})),\nonumber\\
\var(P_s^\mathrm{in})=\frac{1}{\kappa^2}(\var(y_{s-}^\mathrm{out})-\sigma_{in}^2(1-\frac{\kappa^2}{Z^2})),
\label{noiserec}
\end{eqnarray}
where  $\sigma_{in}^2$ is the shot noise of light, assuming that
the incoming light is in a coherent state. The coupling constant
is defined as $\kappa=Z\sqrt{1-e^{-2\gamma_s T}}$. The normalized
EPR variance of atomic noise is
$\xi=\var(P_c^\mathrm{in})+\var(P_s^\mathrm{in})=\var(P_I^\mathrm{in}+P_{II}^\mathrm{in})/2+\var(X_I^\mathrm{in}-X_{II}^\mathrm{in})/2$.
\par

\subsection{Including decay and detection efficiency}

To include the decay of the atomic spin due to spontaneous
emission, magnetic field instabilities, etc. into the input-output
equations (Eq.~\ref{io-eq}), we assume a decay with the rate
$\gamma_\mathrm{extra}$ towards the CSS. Since the interaction
times which are used to perform the read out are short, this is an
adequate approximation.  The atomic part of the input-output
equations becomes
\begin{eqnarray}
P_{c,s}^\mathrm{out}\!=\!P_{c,s}^\mathrm{in}\cdot e^{-\gamma T}
\!-\!\frac{\kappa}{Z^2}{y}_{c,s+}^{\mathrm{in}}\!+\!
\epsilon\sqrt{1\!-\!e^{-2\gamma T}}\cdot{F}_{p,c/s,+},\nonumber\\
X_{c,s}^\mathrm{out}\!=\!X_{c,s}^\mathrm{in}\cdot e^{-\gamma T}
\!+\!\kappa {q}_{c,s+}^{\mathrm{in}}+
\epsilon\sqrt{1\!-\!e^{-2\gamma T}}\cdot{F}_{x,c/s,+}\nonumber,\\
\label{noise}
\end{eqnarray}
with the two-cell noise operators $ F_{i+}=1/N_F\int_0^T
e^{-\gamma(T-t)}  F_i(t) dt$ and
$\langle{F}_{i}^2\rangle=\frac{1}{2}$ and
$\epsilon^2=\gamma_\mathrm{extra}/\gamma$. The relevant light mode
is exponentially growing with the total decay rate
$\gamma=1/T_2=\gamma_s+\gamma_\mathrm{extra}$. The coupling
constant is reduced due to the decay and defined as:
$\kappa=Z\sqrt{(1-\epsilon^2)(1-e^{-2\gamma T})}$. \\
Also the equations for the light are adjusted accordingly:
\begin{eqnarray}
{y}_{c,s-}^{\mathrm{out}}\!\!\!&=&\!\!\!\epsilon^2{y}_{c,s-}^{\mathrm{in}}\!+\!{y}_{c,s+}^{\mathrm{in}}\sqrt{1\!-\!\kappa^2/Z^2}(1\!-\!\epsilon^2)\!+\!\kappa\sqrt{1\!-\!\epsilon^2} {P}_{c,s}^{\mathrm{in}}\nonumber\\
\!\!&+&\!\!\epsilon\sqrt{1-\epsilon^2}Z({F}_{p-}-\sqrt{1-\kappa^2/Z^2}{F}_{p+}),\nonumber\\
{q}_{c,s-}^{\mathrm{out}}\!\!\!&=&\!\!\!\epsilon^2{q}_{c,s-}^{\mathrm{in}}\!+\!{q}_{c,s+}^{\mathrm{in}}\sqrt{1\!-\!\kappa^2/Z^2}(1\!-\!\epsilon^2)\!-\!\kappa/Z^2\!\sqrt{1\!-\!\epsilon^2}{X}_{c,s}^{\mathrm{in}}\nonumber\\
\!\!&+&\!\!\epsilon\sqrt{1\!-\!\epsilon^2}Z({F}_{x-}\!-\!\sqrt{1\!-\!\kappa^2/Z^2}{F}_{x+}).\label{chHO2cellDec}
\end{eqnarray}
The non orthogonal exponentially growing and falling light modes
are now mixed due to the decay.

The reconstruction equation (Eq.~\ref{noiserec}) must then be
adjusted accordingly:
\begin{eqnarray}
\var(P_c^\mathrm{in})=\frac{1}{\kappa^2}(\var(y_{c-}^\mathrm{out})-U^2\cdot\sigma_{s,in}^2-V^2\langle{F}_{i}^2\rangle),\nonumber\\
\var(X_s^\mathrm{in})=\frac{1}{\kappa^2}(\var(y_{s-}^\mathrm{out})-U^2\cdot\sigma_{c,in}^2-V^2\langle{F}_{i}^2\rangle),\label{noiserec2}
 \end{eqnarray}
 with the corrected, reduced coupling
constant and  $U^2(\kappa^2,T_2)$ and $V^2(\kappa^2,T_2)$ which
can be calculated directly from
Eq.~(\ref{chHO2cellDec})~\cite{FootnoteExp1}. To extract the
atomic noise from the light noise measurements, it is therefore
only necessary to know the coupling constant $\kappa$, for which a
measurement procedure is explained below and the decay time $T_2$
which can also easily be measured.
\par
Additionally, the detection efficiency $\eta=0.84(4)$ which arises
from light losses and unperfect detection can be included by
assuming a beam-splitter with transmission $\eta$.

\subsection{Measurement of the coupling strength}\label{chEcoupl}
The most important experimental parameter for the reconstruction
is the coupling constant $\kappa$. It would be possible to
determine $\kappa^2$, by performing noise measurements on known
atomic states, e.g. the CSS or the thermal atomic state. However,
imperfect state preparation or additional noise sources can spoil
such measurements. The approach we implement here is therefore
based on measurements of mean values as opposed to noise
measurement. The modus operandi is to transfer a coherent light
state with a known displacement to the atoms and then read out the
atomic state \cite{WJKRBP10,JWK10}. \par First, a pulse is sent
through two oppositely oriented atomic samples with a displacement
in ${q}_{c,s}^{1st}$, so in $S_z$. Following Eq.\ref{io-eq} and
assuming that the atoms possess no initial displacement, this
leaves the atomic sample with a mean value in the $X$-quadratures.
\begin{equation}
\langle {X}_{c,s}^\mathrm{out} \rangle=\kappa \langle
{q}_{c,s+}^{1st} \rangle.
\end{equation}
To be able to read out those atomic mean values via a measurement
on ${y}_{c,s}$ and thus gain information on $\kappa$, we apply a
$\pi/2$-pulse to the atomic spin rotating ${X}$ into ${P}$. This
can be done by adding a magnetic field in the $\hat x$-direction,
so that the spins rotate a little faster or slower in between the
pulses and $J_{z,I}\rightarrow J_{y,I}$ and $J_{y,II}\rightarrow
J_{z,II}$, leading to $X_{c,s}\rightarrow P_{c,s}$. Then we send a
second light pulse for the read out. The outcome of the light
measurement of the second pulse reveals
\begin{equation}
\langle {y}_{c,s-}^{2nd} \rangle=\kappa \langle
{X}_{c,s}^\mathrm{out} \rangle=\kappa^2\langle
{q}_{c,s+}^{1st}\rangle.
\end{equation}
The coupling strength can be calculated:
$\kappa^2=\frac{\avg{{y}_{c,s-}^{2nd}}}{\avg{{q}_{c,s+}^{1st}}}$.

\par The displaced coherent light states are produced with the help of an electro optical modulator (EOM) \cite{Sh06}. The strongly polarized beam is
sent through an EOM whose optical axis is slightly tilted compared
to the input polarization. Then a DC voltage and a small
modulation at 322 kHz can be used to rotate a small portion of the
large polarization component in $\hat{y}$-direction into the
x-polarization mode. The value of the DC voltage determines the
phase of the modulation in phase space, so the position in the
${S}_2$-${S}_3$ plane. The strength and phase of the RF modulation
determine the size of the displacement of the cosine and sine
modes.
\par
For the measurement of $\avg{q_{c,s}^{1st}}$, a
$\frac{\lambda}{4}$-plate is inserted in the detection path to
switch to a $ S_3$ measurement~\cite{FootnoteExp2}.\par

\begin{figure}
   \centering
\includegraphics[width=\columnwidth]{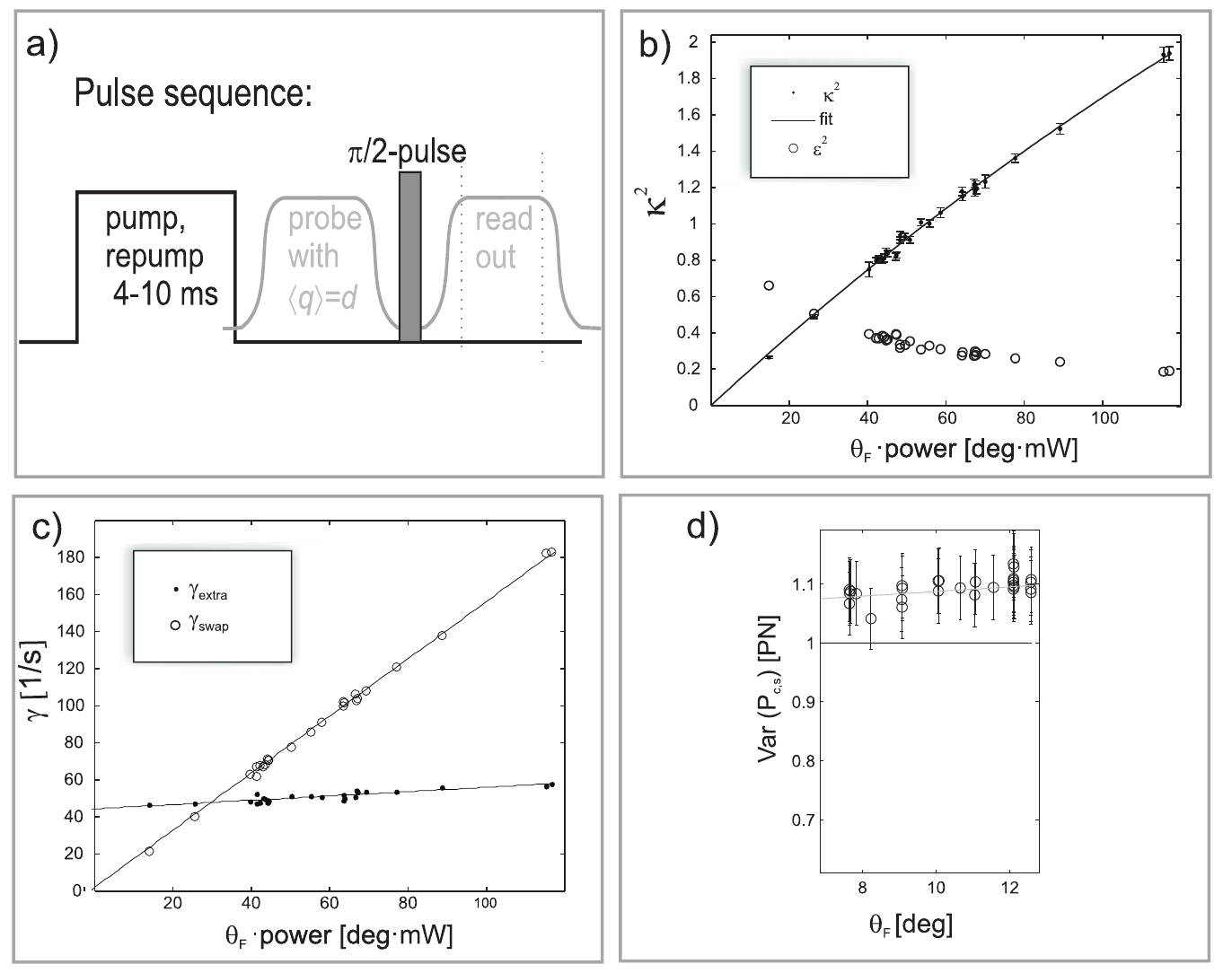}
   \caption[]{In a) the pulse sequence is shown. The probe light is turned on and off smoothly to avoid noise contributions at 322kHz.
   b) shows $\kappa^2$ for a varied number of atoms and a light power $P=5mW$, in
   c) the corresponding decay constants $\gamma_\mathrm{extra}$ and $\gamma_{s}$ are shown. In d) the scaling of the normalized atomic noise over $N$ is shown.}\label{chEkappa}
\end{figure}

In Fig.~\ref{chEkappa} measurements of $\kappa^2$ are shown for
different numbers of atoms $N$ with a fixed probe power of $5$mW
and a probe duration of ~1ms. $N$ is varied by changing the
temperature of the atomic ensemble. It can be monitored by sending
a weak linearly polarized probe beam in the direction of the
macroscopic orientation. Due to the Faraday effect the light
polarization is rotated by $\theta\propto J_x$ and for the CSS
$J_x\approx4\cdot N$. The orientation of the atomic ensembles can
be tested via magneto optical resonance spectroscopy
\cite{Julsgaard2004}. Orientations of $0.997(3)$ are reached
regularly in this experimental setup. Fig.~\ref{chEkappa}c
displays how the measured decay rate $\gamma=1/T_2$ can be
decomposed in $\gamma_{s}$ and $\gamma_{\mathrm{extra}}$. Clearly
$\gamma_{s}\propto P\cdot\theta_F$.

\subsection{PN measurement}
When the coupling constant is known, Eq.~(\ref{noiserec2}) can be
used to reconstruct the collective atomic operator noise from
measurements of noise on $y_{c,s-}$ of the outgoing light. In
Fig.~\ref{chEkappa}d measurements of the atomic noise in units of
projection noise (PN) are shown for different $N$. There is a
small additional noise component, probably arising from technical
noise, which scales with the number of atoms. Only a small range
of $\theta_F$ and thus $N$ is shown. For higher $N$ additional
classical noise of unknown origin is measured, disqualifying
higher atomic densities as a working point for quantum noise
limited measurements in this setup.

\par
To find the time evolution of the atomic noise, while probe light
is present (as the  curves shown in
Fig.~\ref{Fig:PureDissipation}a), $y_{c,s-}$ is evaluated in the
time-interval $[t,t+t_\mathrm{probe}]$, where $t_\mathrm{probe}$
is the evaluation time with $\kappa(t_\mathrm{probe})$, to find
the atomic variances at time $t$. The corresponding pulse sequence
is shown on the left of Fig.~\ref{pulseseq}.

\begin{figure}
   \centering
   \includegraphics[width=6cm]{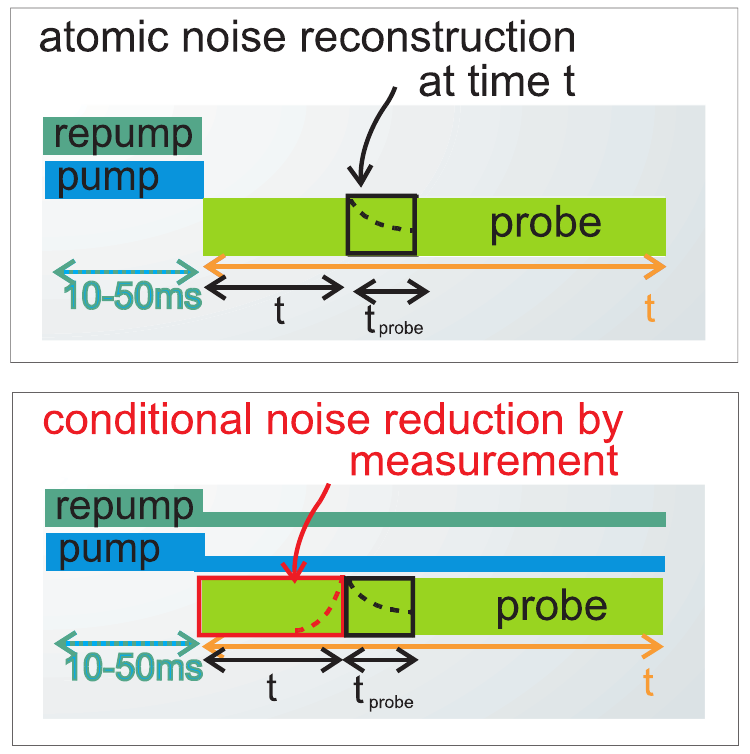}\label{pulseseq}
   \caption[]{Pulse sequences for long probe durations. On the left a pulse sequence with a long probe pulse is depicted, displaying the evaluation time interval $[t;t_\mathrm{probe}]$ used to find the atomic noise at time $t$. On the right it is shown how a measurement prior to $t$ can be used to conditionally reduce the atomic noise. For measurements like this typically a fast exponentially growing mode function is used.}\label{pulseseq}
\end{figure}

\subsection{Measurement of the conditional atomic noise variance}

The theoretical background for generation of the steady state of
atoms with the noise reduced by combining dissipation with the
measurement is described in
Sec.~\ref{Sec:EntanglementAssistedByMeasurements}. The
experimental procedure for the generation of a state with reduced
conditional noise variance is described in the following. We wish
to squeeze the atomic variances at time $t$ by measuring the
outgoing light operators $y_{c}$ and $ y_{s}$ for the preceeding
time. The time sequence is illustrated on the right of
Fig.~\ref{pulseseq}. The probe pulse is divided in sections, where
measurements in the first time interval $[0,t]$ are used for the
conditional noise reduction of the atomic operators at time $t$,
and the consecutive time slice $[t, t+ t_{probe}]$ is used for the
reconstruction of the atomic state as described above. Here, the
quantity of interest is the conditionally reduced atomic variance
at time $t$:
$\mathrm{var}(P^\mathrm{out}_{c/s})\!_{\mathrm{\tiny{con}}}=\mathrm{var}(P^{\mathrm{out}}_{c/s}-\alpha
y^\mathrm{probe}_{c/s})= \mathrm{var}(\frac{1}{\sqrt{2J_x}}(
J_{y,z,I} \!\pm\!
J_{y,z,II})-\alpha\frac{1}{\sqrt{2\phi^\mathrm{probe}}}{S}_{2c/s}^{\mathrm{probe}})$
(compare Eq.~\ref{ConditionalVariance}). The superscript "probe"
refers to the first time slice from 0 to $t$. To achieve an
optimal noise reduction, $\alpha$ and the temporal mode function
of the probe section are optimized. The optimized mode function is
exponentially rising with a rate that is typically faster than
$1/T_2$. This means that only the last bit of the long preceding
pulse is used for the conditional noise reduction. This behavior
is rooted in the additional decoherence mechanisms. The atomic
state acquires a noise component piling up with rate
$\gamma_\mathrm{extra}$. The measurements closest in time to the
atomic state which one wishes to squeeze, should be weighted most.
\par The actual measurement of $\mathrm{var}(P^\mathrm{out}_{c/s})_{\mathrm{\tiny{con}}}$ is done by evaluating  $y_{c/s}^\mathrm{out}$ in the second time slice $[t;t+t_{probe}]$. The same reconstruction mechanism as for the unconditional atomic state reconstruction is used, therefore  is the second time slice
evaluated with an exponentially falling mode function with
$\gamma=\gamma_s+\gamma_\mathrm{extra}$. Again $\var
(y_{c/s-}^\mathrm{out})$ is utilized to establish the value of
$\mathrm{var}(P^\mathrm{out}_{c/s})$ at time $t$.
  Accordingly, $\mathrm{var}(y^\mathrm{out}_{c/s})_{\mathrm{\tiny{con}}}=\mathrm{var}(y_{c/s-}^\mathrm{out}- \alpha^*{y}_{c/s}^{\mathrm{probe}})$
can be used to find
$\mathrm{var}(P^\mathrm{out}_{c/s})_{\mathrm{\tiny{con}}}$ with
the same reconstruction procedure. The conditional variance can be
extracted from the two-time correlation functions
  $\avg{{y}_{c/s}^{\mathrm{probe}}{y}_{c/s-}^\mathrm{out}}$:
$\sigma_{cond,out}^2=\mathrm{var}(y_{c/s-}^\mathrm{out})+{\alpha^*}^2\mathrm{var}(y_{c/s}^{\mathrm{probe}})-
2\alpha^* \avg{{y}_{c/s}^{\mathrm{probe}}{y}_{c/s-}^\mathrm{out}}$
by optimizing $\alpha^*$, this is assuming that there are no
correlations between the light operators at different points in
time $<y(t)y(t')>=\delta(t-t')$.

\section{Conclusions}\label{Sec:Conclusions}

The method discussed in this article allows one to
deterministically generate entangled states between two atomic
ensembles over a macroscopic distance. The desired quantum state
is thereby stabilized by the dissipative mechanism, which renders
the created entanglement robust and long-lived. Purely dissipative
entanglement generation does not require postselection nor
conditioning.

For two-level systems, the dissipative mechanism explained in
Sec.~\ref{Subsec:PurelyDissipativeEntanglement} can be used to
produce steady state entanglement, i.e. entanglement which is
available permanently. This is even possible in the presence of
imperfections and noise sources. The use of atoms with a
multi-level ground state complicates the generation of steady
state entanglement since atoms can undergo transitions to internal
levels, which are not coupled to the engineered dynamics in the
desired way. Atoms with a two-level ground state such as Ytterbium
($^{171}$Yb) appear therefore best suited for the realization of
the discussed scheme. However, steady state entanglement can also
be created in ensembles of atoms with a multilevel structure if
strong pumping fields are applied, which transfer the atoms
(incoherently) back to a specific two-level sub-system. However,
this strategy requires a high optical depth of the atomic sample,
or the use of a low finesse cavity. Therefore, a hybrid solution
has been implemented in~\cite{EbDExperiment}, which combines the
dissipative mechanism with measurements. As explained in
Sec.~\ref{Sec:EntanglementAssistedByMeasurements}, measurements on
the light field yield information on the atomic state and can
therefore be used to obtain a quantum state of higher purity and
accordingly produce higher entangled states.

The mechanism considered here is applicable in any system which
can be coupled to two photonic sideband modes according through a
tunable quadratic interaction, for example for the generation of
entanglement between two mechanical oscillators. Even though the
limited coherence times of quantum systems impose typically severe
restrictions on the life times of quantum superposition states,
dissipative methods for quantum state engineering allow one to
produce event-ready quantum states for applications in Quantum
Information Science. Exploring and exploiting all advantages of
dissipative approaches will require both, devising new protocols
which are capable of generating and processing steady states as
well as finding realistic and practical ways of implementing the
required coupling of physical systems to a bath.

\subsection*{Acknowledgements}

We acknowledge funding by DARPA through the QuSAR program and the
EU projects MALICIA, Q-ESSENCE and QUEVADIS. C.A.M. acknowledges
support from the Alexander von Humboldt Foundation and the Elite
Network of Bavaria (ENB) project QCCC.

\appendix

\section{Derivation of the master equation}\label{App:MasterEquation}
The interaction of atoms and light illustrated in
Fig.~\ref{Fig:Setup} can be described by the effective ground
state Hamiltonian
\begin{eqnarray*}
H=H_{\mathrm{A}}+H_{\mathrm{L}}+H_{\mathrm{int}},
\end{eqnarray*}
where excited states have been eliminated using the fact that the
detuning $|\Delta|$ is large compared to the Doppler width
$\delta_{\mathrm{Doppler}}$ and atomic decay rates
$\Gamma_{\mathrm{atomic}}$. Here and in the following,
$\Gamma_{\mathrm{atomic}}$ denotes the largest effective rate for
atomic ground states including single particle as well as
collective rates (see below).
$H_{\mathrm{A}}=\Omega \left(J_{x,I}-J_{x,II}\right)$ accounts for
the Zeeman splitting of the atoms in the external magnetic field
and $H_{\mathrm{L}}=\int dk\ \omega_k \
a_{\mathbf{k}}^{\dag}a_{\mathbf{k}}$ is the free Hamiltonian of
the light field, where $a_\mathbf{k}$ is the annihilation operator
of a photon with wave vector $\mathbf{k}$ and frequency
$\omega_{k}$. In a rotating frame, the interaction Hamiltonian is
given by
\begin{eqnarray}
H_{\mathrm{int}}\!\!\!&=&\!\!\!\!\!\int_{\Delta
\omega_{ls}}\!\!\!\!\!\! \!\!d
\mathbf{k}\!\sum_{\lambda_{\mathbf{k}}}\!
g(\mathbf{k})\!\!\!\left(\!\!\mu\!\sum_{i=1}^{N}\!\sigma_{I,i}e^{i\Delta\mathbf{k}
\mathbf{r}_{i}}
\!\!-\!\!\nu\!\sum_{j=1}^{N}\!\sigma_{II,j}e^{i\Delta\mathbf{k}\
\mathbf{r}_{j}}\!\right)\!a_{\mathbf{k}}^{\dag}\nonumber\\
\!\!\!&+&\!\!\!\!\!\int_{\Delta \omega_{us}}\!\!\!\!\!\!\!\!d
\mathbf{k}\!\sum_{\lambda_{\mathbf{k}}}\! g(\mathbf{k})\!\!\!
\left(\!\!\mu\!\sum_{i=1}^{N}\!\sigma_{II,i}^{\dag}e^{i\Delta\mathbf{k}
\mathbf{r}_{i}}
\!\!-\!\!\nu\!\sum_{j=1}^{N}\!\sigma_{I,j}^{\dag}e^{i\Delta\mathbf{k}\
\mathbf{r}_{j}}\!\right)\!a_{\mathbf{k}}^{\dag} \nonumber\\
\!\!\!&+&\!\!\!H.C.\ ,\label{Hamiltonian}
\end{eqnarray}
where $\lambda_{\mathbf{k}}$ specifies the the two orthogonal
polarizations of the light mode with wave vector $\mathbf{k}$. The
first and second integral cover narrow bandwidths $\Delta
\omega_{ls}$ and $\Delta \omega_{us}$ centered around the lower
and upper sideband respectively. The atomic operator
$\sigma_{I/II,i}=|\!\!\uparrow\rangle_{I/II,i}\langle \downarrow
\!\!|$ refers to a particle in ensemble $I/II$ at position
$\mathbf{r}_{i}$, $\Delta\mathbf{k}=\mathbf{k}_{L}-\mathbf{k}$ and
$\mathbf{k}_{\mathrm{L}}$ is the wavevector of the applied
classical field. $g(\mathbf{k})\mu$ and $g(\mathbf{k})\nu$ denote
the effective coupling strengths of the passive
(beamsplitter-like) part of the interaction and the active
(squeezing) component of the Hamiltonian respectively. The laser
field covers a very narrow bandwidth around the central frequency
$\omega_{\mathrm{L}}$ and is sufficiently off-resonant such that
the interaction is well within the dispersive regime and
absorption effects can be
neglected.\\
\\Starting from Hamiltonian (\ref{Hamiltonian}), a master equation of
Lindblad form can be derived for the reduced atomic density matrix
$\rho(t)$. To this end, the Born Markov approximation is used,
which is well justified for optical frequencies. Since the Larmor
splitting exceeds well atomic decay rates
$\Omega>>\Gamma_{\mathrm{atomic}}$, the two sideband modes can be
treated as independent baths.
The effect of atomic motion gives rise to noise terms and can be
included in the master equation in the form of averaged
coefficients, where the average in time corresponds to an average
in space. This is legitimate in the fast motion limit, where the
time scale set by the average velocity of the atoms $v$ is fast
compared to the time scale of the radiative decay
$\Gamma_{\mathrm{atomic}} \frac{L}{v} \ll 1$. In this case, the
emission of light can be described
independently of the evolution of atomic positions.\\
\\Using the definitions
\begin{eqnarray*}
  A&=&\mu \frac{1}{\sqrt{N}}\sum_{i=1}^{N}\sigma_{I,i}-\nu \frac{1}{\sqrt{N}}\sum_{i=1}^{N}\sigma_{II,i},\\
  B&=&\mu \frac{1}{\sqrt{N}}\sum_{i=1}^{N}\sigma_{II,i}^{\dag}-\nu \frac{1}{\sqrt{N}}\sum_{i=1}^{N}\sigma^{\dag}_{I,i},\nonumber
\end{eqnarray*}
the resulting master equation can be cast in the form
\begin{eqnarray}\label{AtomicMotion}
\frac{d}{dt}\ \! \rho(t)\!\!&=&\!\!\frac{1}{2}d\Gamma  A \rho(t)
A^{\dag} \!+\! \frac{1}{2}d\Gamma  B \rho(t) B^{\dag}\nonumber\\
\!\!&+&\!\!\frac{1}{2}\Gamma\mu^2\sum_{i=1}^{N}\left(
\sigma_{I,i}\rho(t)\sigma_{I,i}^{\dag}\!+\!
\sigma_{II,i}^{\dag}\rho(t)\sigma_{II,i}\right)\nonumber\\
\!\!&+&\!\!\frac{1}{2}\Gamma\nu^2\sum_{i=1}^{N}\left(
\sigma_{I,i}^{\dag}\rho(t)\sigma_{I,i}\!+\!
\sigma_{II,i}\rho(t)\sigma_{II,i}^{\dag}\right)\nonumber\\
\!\!+\!\!...\ ,
\end{eqnarray}
where $\Gamma$ is the single particle decay rate and a short hand
notation was used. Master equations of Lindblad form $\frac{d
}{dt}\rho(t)=\frac{\gamma}{2} \left(A
\rho(t)A^{\dag}-A^{\dag}A\rho(t)\right)+H.C.$ with decay rate
$\gamma$ and jump operator $A$ are abbreviated by the expression
$\frac{d }{dt}\rho(t)=\frac{\gamma}{2} A \rho(t)A^{\dag}+...\ $.
$d$ denotes the resonant optical depth of one atomic ensemble.
Note that the entangling terms in the first line are enhanced by a
factor $d$, such that for sufficiently optically thick samples,
the additional noise terms reflecting thermal motion are small
compared to the
desired contributions.\\
\\Next, additional cooling and heating processes, as well as
dephasing are included. The full master equation is given by
\begin{eqnarray}\label{FullME}
\frac{d}{dt}\ \!\rho(t)\!\!&=&\!\! \frac{1}{2}d\Gamma  A \rho(t)
A^{\dag} \!+\! \frac{1}{2}d\Gamma  B \rho(t) B^{\dag}\\
\!\!&+&\!\!\frac{1}{2}\Gamma_{\text{\tiny{cool}}}\sum_{i=1}^{N}\left(
\sigma_{I,i}\rho(t)\sigma_{I,i}^{\dag}+
\sigma_{II,i}^{\dag}\rho(t)\right)\nonumber\\
\!\!&+&\!\!\frac{1}{2}\Gamma_{\text{\tiny{heat}}}\sum_{i=1}^{N}\left(
\sigma_{I,i}^{\dag}\rho(t)\sigma_{I,i}\!+\!
\sigma_{II,i}\rho(t)\sigma_{II,i}^{\dag}\right)\nonumber\\
\!\!&+&\!\!\frac{1}{2}\Gamma_{\rm{\text{deph}}}\!\sum_{i=1}^{N}\!\!\left(\sigma_{\downarrow\downarrow,I,i}\rho(t)
\sigma_{\downarrow\downarrow,I,i} \!+\!
\sigma_{\downarrow\downarrow,II,i}\rho(t)
\sigma_{\downarrow\downarrow,II,i}\!\right),\nonumber\\
\!\!&+&\!\!...\ \nonumber,
\end{eqnarray}
with
$\sigma_{\uparrow\uparrow,I/II,i}=|\!\!\uparrow\rangle_{I/II,i}\langle
\uparrow\!\!|$ and
$\sigma_{\downarrow\downarrow,I/II,i}=|\!\!\downarrow\rangle_{I/II,i}\langle
\downarrow\!\!|$. The noise terms proportional to $\Gamma\mu^2$
and $\Gamma\nu^2$ in Eq.~(\ref{AtomicMotion}) have been absorbed
in the second and third line of Eq.~(\ref{FullME}) respectively.
The last three lines represent single particle processes. Hence
they do not feature a collective enhancement factor as the
entangling terms in the first line.
As shown in \cite{EbDtheory}, Eq.~(\ref{FullME}) includes all
terms that need to be taken into account. Collective dephasing
terms, as well as collective contributions due to pump and repump
fields can be neglected. Similarly, the distance $R$ between the
two ensembles does not play a role for $k_L\gg R/L^2$ and $k_L
L\gg1$, where $L$ is the spatial extend of an atomic ensemble. For
the experimental setting under consideration this condition is
fulfilled.
\section{Calculation of entanglement}\label{App:CalculationEntanglement}
Below it is shown how the entanglement measured by the quantity
$\xi=\Sigma_{J}/\left(2|\langle J_x\rangle|\right)$, where
$\Sigma_{J}=\mathrm{var} (J_{y,\mathrm{I}}- J_{y,\mathrm{II}})+
\mathrm{var}(J_{\mathrm{z},I} - J_{z,\mathrm{II}})$,
can be determined in the limit $N\gg 1$, assuming that the number
of atoms in the two-level system $N_2$ depends on time.
For clarity, operators referring to the two-level model are
labelled with subscript "2".
The time derivative of the variance $\Sigma_{J_2}$ is calculated
using Eq.~(\ref{FullME}). By applying the decorrelation
approximation $\langle J_{y/z} (t)J_x(t)\rangle_2\approx\langle
J_{y/z}(t) \rangle_2\langle J_x(t) \rangle_2$ for mean values of
products of transverse and longitudinal spins one obtains
\begin{eqnarray*}
\frac{d }{dt}\ \!\Sigma_{J_2}(t)
&=&-\left(\tilde{\Gamma}+d(t)\Gamma P_2(t)\right)\Sigma_{J_2}(t)
\\
&+&N_2(t)\left(\tilde{\Gamma}+d(t)\Gamma
P_2(t)^2\left(\mu-\nu\right)^2\right),
\end{eqnarray*}
where $d(t)=d\ N_{2}(t)/N$,
$\tilde{\Gamma}=\Gamma_{\rm{cool}}+\Gamma_{\rm{heat}}+\Gamma_{\rm{deph}}$
and $P_2(t)=2\langle J_x(t)\rangle/\left(N_2(t)\right)$. For
$t\rightarrow\infty$ and $N_2=N$,
\begin{eqnarray*}
\Sigma_{J_{2,\infty}}&=&\ N\ \frac{\tilde{\Gamma}+d\Gamma
P_{2,\infty}^2\left(\mu-\nu\right)^2}{\tilde{\Gamma}+d\Gamma
P_{2,\infty}}\ .
\end{eqnarray*}
Next, the time evolution of the longitudinal spin is considered.
Eq.~(\ref{FullME}) yields
\begin{eqnarray*}
\frac{d}{dt}\ \!\langle
J_x(t)\rangle_2&=&-\left(\Gamma_{\rm{heat}}+\Gamma_{\rm{cool}}\right)\langle
J_x(t)\rangle_2\\
&+&\frac{N_2(t)}{2}\left(\Gamma_{\rm{cool}}-\Gamma_{\rm{heat}}\right),
\end{eqnarray*}
such that for constant particle number $N_2=N$,
\begin{eqnarray*}
\langle J_x\rangle_{2,\infty}=\frac{N}{2}\
\frac{\Gamma_{\rm{cool}}-\Gamma_{\rm{heat}}}{\Gamma_{\rm{cool}}+\Gamma_{\rm{heat}}},
\end{eqnarray*}
and therefore
\begin{eqnarray}\label{Eq:SteadyState}
 \xi_{2,\infty}&=&\
 \frac{1}{P_{2,\infty}}\ \frac{\tilde{\Gamma}+d\Gamma
 P_{2,\infty}^2\left(\mu-\nu\right)^2}{\tilde{\Gamma}+d\Gamma
 P_{2,\infty}},\\
 P_{2,\infty}&=&\frac{\Gamma_{\rm{cool}}-\Gamma_{\rm{heat}}}{\Gamma_{\rm{cool}}+\Gamma_{\rm{heat}}}
\end{eqnarray}
in the steady state. In the limit $d\rightarrow\infty$, this
equation reduces to $
\xi_{2,\infty}=\left(\mu-\nu\right)^2$.\\
\\The variation of $N_2(t)$ and $P_2(t)$ is slow
compared to the evolution of $\Sigma_{J_{2}}(t)$. In the limit
where the entangled quantum state follows the changing particle
number and atomic polarization adiabatically, $\xi_2(t)$ is given
by
\begin{eqnarray}\label{EPR}
\xi_2(t)&=&\frac{\Sigma_{J_2}(0)}{2P_2(t)}\
\!e^{-\left(\tilde{\Gamma}+d(t)\Gamma P_2(t)\right)t}\\
&+&\frac{\tilde{\Gamma}+d(t)\Gamma
P_2(t)^2(\mu-\nu)^2}{P_2(t)\left(\tilde{\Gamma}+d(t)\Gamma
P_2(t)\right)}\left(1-e^{-\left(\tilde{\Gamma}+d(t)\Gamma
P_2(t)\right)t}\right).\nonumber
\end{eqnarray}
\section{Towards purely dissipative steady state entanglement using incoherent pump fields}\label{App:FitData}
In the following, we explain how the estimates shown in
Fig.~\ref{Fig:PureDissipation}b are obtained. We use here a
simplified model, which has been employed in~\cite{EbDExperiment}
to compare the measured results to the theoretical predictions.
Details regarding these fits can be found in the Supplemental
Material of~\cite{EbDExperiment}. We model the experiment in terms
of three atomic levels $|\!\!\uparrow\rangle_I\equiv|4,
4\rangle_I$, $|\!\!\downarrow\rangle_I\equiv|4,3\rangle_{I}$ and
$|h\rangle_{I}\equiv|3, 3\rangle_I$
($|\!\!\uparrow\rangle_{II}\equiv|4, -3\rangle_{II}$,
$|\!\!\downarrow\rangle_{II}\equiv|4,-4\rangle_{II}$ and
$|h\rangle_{II}\equiv|3, -3\rangle_{II}$) for the first (second)
ensemble. For the timescales considered here, the atomic
population in other internal states is negligible.
In order to describe the underlaying physics qualitatively, using
only a small number of parameters, we assume further that
$\Gamma_{|4,\pm4\rangle \rightarrow|h\rangle}\approx
\Gamma_{|4,\pm3\rangle \rightarrow|h\rangle}=\Gamma_{\rm{out}}$
and $\Gamma_{|h\rangle\rightarrow |4,\pm4\rangle}\approx
\Gamma_{|h\rangle \rightarrow|4,\pm3\rangle}=\Gamma_{\rm{in}}$,
where the abbreviations $\Gamma_{|4,\pm 4\rangle\rightarrow|4,\pm
3\rangle}=\Gamma_{4,3}$ and $\Gamma_{|4,\pm
3\rangle\rightarrow|3,\pm 4\rangle}=\Gamma_{3,4}$ have been
used.\\
Atomic transitions are taken into account by introducing the
collisional rate $\Gamma_{\mathrm{col}}$. Since the atomic thermal
energy is large compared to the level splittings, we assume the
same rate $\Gamma_{\mathrm{col}}$ for all atomic transitions.
Finally, we include $\sigma_{\pm}$ polarized pump and repump
fields, which induce resonant transitions with $\Delta m_{F}\pm 1$
in the first/second ensemble as shown in Fig.~\ref{Fig:Exp}b. Pump
fields drive transitions within the manifold of atomic states with
$F=4$ and repump fields transfer atoms from states with $F=3$ back
to $F=4$. In this case,
\begin{eqnarray*}
\Gamma_{\mathrm{out}}&=&\Gamma_{\mathrm{L}}^{\mathrm{out}}+\Gamma_{\mathrm{col}},\\
\Gamma_{\mathrm{in}}&=&\Gamma_{\mathrm{repump}}+\Gamma_{\mathrm{col}}.
\end{eqnarray*}
where $\Gamma_{\mathrm{L}}^{\mathrm{out}}$ is the rate at which
atoms leave the two-level subsystem due to radiative transitions
caused by the driving field. $\Gamma_{\mathrm{repump}}$ is the
rate at which the applied repump fields transfer atoms back.
Transitions within the two-level subsystem occur at the rates
\begin{eqnarray*}
\Gamma_{3,4}&=&\mu^2\ \Gamma+\Gamma_{\mathrm{pump}}+\Gamma_{\mathrm{col}},\\
\Gamma_{4,3}&=&\nu^2\ \Gamma+\Gamma_{\mathrm{col}}.
\end{eqnarray*}
$\mu^2\Gamma$ and $\Gamma_{\mathrm{pump}}$ are the driving field
and pump field induced cooling rates respectively. The heating
rate caused by the driving field is given by $\nu^2\ \Gamma$. Note
that the application of pump fields leads to an increased
dephasing rate $\tilde{\Gamma}$. In contrast, repump fields do not
have an effect on $\tilde{\Gamma}$~\cite{EbDtheory}. We estimate
the effect of these fields on the dephasing rate by adding
$2\Gamma_{\mathrm{pump}}$ to $\tilde{\Gamma}$.\\
\\Fig.~\ref{Fig:PureDissipation}b shows the predicted time
evolution of entanglement for $d=55;100;150$. The parameters take
the values used to fit the experimental data measured in the
absence of additional fields (compare Fig.~2, panels a and b in
\cite{EbDExperiment}), $\Gamma=0.002$ms$^{-1}$,
$\tilde{\Gamma}=0.193$ms$^{-1}$,
$\Gamma_{\mathrm{col}}=0.002$ms$^{-1}$ and $Z=2.5$. The presence
of both, pump and repump fields is included as described above
with
$\Gamma_{\mathrm{pump}}=\Gamma_{\mathrm{repump}}=0.160$ms$^{-1}$.

%
%
%

%

\begin{thebibliography}{10}
\bibitem{FootnoteMacroscopicity}
Note that the term "entanglement between macroscopic objects" does
not refer to a macroscopic superposition since the generated
entanglement corresponds to less than one e-bit.
\bibitem{FootnoteEPR}
In the famous example introduced by Einstein, Podolski and Rosen,
the quantum state under consideration is a simultaneous eigenstate
of the sum of positions $x_{+}=x_{I}+x_{II}$ and difference of
momenta $p_{-}=p_{I}-p_{II}$ of two separate particles. This
corresponds to an infinitely squeezed state with
$\mathrm{var}(x_+)=\mathrm{var}(p_-)=0$, where the
momenta/positions are perfectly correlated/anticorrelated.
\bibitem{FootnoteCesium}
$^{133}$Cs is an alkali atom with nuclear spin $I=3/2$. The ground
state is split into two manifolds with total spin $F=I\pm1/2$.
\bibitem{FootnoteMeasurement}
A measurement yielding the result e, produces the quantum state
$\rho_{e}$. The situation where no measurement is performed
corresponds to the situation where the measurement result is
unknown. The resulting mixed state is given by
$\tilde{\rho}=\sum_e P(e) \rho_{e}$, where $P(e)$ is the
probability to obtain the measurement result e and the sum
$\sum_e$ runs over all possible measurement results.
\bibitem{FootnoteRemark}
In the experiment reported on here, the light field is measured
continuously due to technical reasons. However, the measurement
results are not used in the first set of experiments, where
entanglement is created purely dissipatively. This corresponds to
a situation where the measurement outcomes are not known or,
equivalently, no measurement is performed.
\bibitem{FootnoteNoise}
In this case the Hamiltonian is given by $H=H_{c}+H_{s}$, where
$H_{c/s}=H^{\mathrm{{ent}}}_{c/s}+H^{\mathrm{{noise}}}_{c/s}$.
$H^{\mathrm{{ent}}}_{c/s}=\sqrt{2\gamma_s}\left(\frac{1}{Z}
X_{c/s}y_{c/s}+Z P_{c}q_{c/s}\right)$ corresponds to the desired
entangling interaction, while
$H^{\mathrm{{noise}}}_{c/s}=\sqrt{2\gamma_{\mathrm{extra}}}\left(X_{c/s}x_{N,c/s}+P_{c/s}p_{N,c/s}\right)$
describes transverse dephasing at a rate
$\gamma_{\mathrm{extra}}$. $x_{N,c/s}$ and $p_{N,c/s}$ refer to
noise modes with $\langle x_{N,c/s}(t)\rangle=\langle
p_{N,c/s}(t)\rangle=0$, $\langle
x_{N,c/s}(t)x_{N_{c,s}}(t')\rangle=\langle
p_{N,c/s}(t)p_{N_{c,s}}(t')\rangle=\delta(t-t')/2$ and
$[x_{N,c/s},p_{N,s/c}]=0$.
\bibitem{FootnoteExp1}
Note that the exponentially growing and falling modes are not
orthogonal. Taking this into account:
$U^2=\epsilon ^4+e^{-2 T \gamma } \left(-1+\epsilon ^2\right)^2-2 \sqrt{e^{-2 T \gamma }} T \gamma  \epsilon ^2 \left(-1+\epsilon ^2\right)\mathrm{Csch}(T \gamma) $ and \\
$V^2=\frac{e^{-2 T \gamma } \epsilon ^2 (-1+\epsilon ^2)(-1-e^{2 T
\gamma }+2 e^{2 T \gamma } \sqrt{e^{-2 T \gamma }} T \gamma
\mathrm{Csch}[T \gamma ])}{\xi ^2}\approx 0$.
 $\epsilon^2$ can be easily extracted from $\kappa^2$:
$\epsilon^2=(1-\kappa^2)/(1-e^{-2T/T2})/Z^2$
\bibitem{FootnoteExp2}
To test if the measurement is set up correctly, one creates a
fixed signal coming from the atoms in one atomic ensemble e.g.
created by a magnetic RF pulse. Following Eq.~\ref{io-eq}, the
atomic signal should be suppressed by $1/Z^2$ in a $\avg{ S_3}$-
compared to a $\avg{ S_2}$-measurement.
\end{thebibliography}
\end{document}